\documentclass[pdflatex,sn-basic]{sn-jnl}% Basic Springer Nature Reference Style/Chemistry Reference Style

\jyear{2021}%

\theoremstyle{thmstyleone}%
\theoremstyle{thmstyletwo}%

\theoremstyle{thmstylethree}%

\chardef\us=`\_

\newcommand{\helix}[1]{{H\tiny{E}\normalsize{LI}\tiny{X}$^{{\textstyle+}}$#1}} 
\newcommand{\sophi}{{SO/PHI}}
\newcommand{\so}{{SO}}
\newcommand{\instr}{{PHI}}
\graphicspath{{Figures/}} %Setting the graphicspath

\usepackage{bm}
\usepackage{amssymb,enumitem,mathtools}

\newcommand{\ion}[2]{\textup{#1\,\textsc{\lowercase{#2}}}}
\newcommand*\element[1][]{%
  \def\aa@element@tr{#1}%
  \aa@element
}

\begin{document}

\title[Accuracy analysis of the on-board data reduction pipeline for SO/PHI]{Accuracy analysis of the on-board data reduction pipeline for the Polarimetric and Helioseismic Imager on the Solar Orbiter mission}


\author*[1,2]{\fnm{Kinga} \sur{Albert}}\email{albert@mps.mpg.de}

\author[1]{\fnm{Johann} \sur{Hirzberger}}

\author[1,3]{\fnm{J.~Sebasti\'an} \sur{Castellanos~Dur\'an}}

\author[4]{\fnm{David} \sur{Orozco~Su\'arez}}

\author[1]{\fnm{Joachim} \sur{Woch}}

\author[5]{\fnm{Harald} \sur{Michalik}}

\author[1]{\fnm{Sami K.} \sur{Solanki}}

\affil*[1]{\orgname{Max Planck Institute for Solar System Research}, \orgaddress{\street{Justus-von-Liebig-Weg}, \city{G\"ottingen}, \postcode{37077}, \country{Germany}}}

\affil[2]{\orgname{Technnical University of Braunschweig}, \orgaddress{\street{Hans-Sommer-Stra$\ss$e}, \city{Braunschweig}, \postcode{3329}, \country{Germany}}}

\affil[3]{\orgname{Georg-August-Universit\"at G\"ottingen}, \orgaddress{\street{Friedrich-Hund-Platz}, \city{G\"ottingen}, \postcode{37077}, \state{State}, \country{Country}}}

\affil[4]{\orgname{Instituto de Astrof\'isica de Andaluc\'ia (IAA - CSIC)}, \orgaddress{\street{Apartado}, \city{Granada}, \postcode{3004}, \country{Spain}}}

\affil[5]{\orgname{Institute of Computer and Network Engineering at the TU Braunschweig}, \orgaddress{\street{Hans-Sommer-Stra$\ss$e}, \city{Braunschweig}, \postcode{3329}, \country{Germany}}}


%\abstract{}

%%================================%%
%% Sample for structured abstract %%
%%================================%%

\abstract{\textbf{Context:} Scientific data reduction on-board deep space missions is a powerful approach to maximise science return, in the absence of wide telemetry bandwidths. The Polarimetric and Helioseismic Imager (\instr{}) on-board the Solar Orbiter (\so{}) is the first solar spectropolarimeter that opted for this solution, and provides the scientific community with science-ready data directly from orbit. This is the first instance of full solar spectropolarimetric data reduction on a spacecraft.

\textbf{Methods:} In this paper, we analyse the accuracy achieved by the on-board data reduction, which is determined by the trade-offs taken to reduce computational demands and to ensure the autonomous operation of the instrument during the data reduction process. We look at the magnitude and nature of errors introduced in the different pipeline steps of the processing. We use an MHD sunspot simulation to isolate the data processing from other sources of inaccuracy. We process the data set with calibration data obtained from \sophi{} in orbit, and compare results calculated on a representative \sophi{} model on ground with a reference implementation of the same pipeline, without the on-board processing trade-offs.

\textbf{Results:} Our investigation shows that the accuracy in the determination of the Stokes vectors, achieved by the data processing, is at least two orders of magnitude better than what the instrument was designed to achieve as final accuracy. Therefore, the data accuracy and the polarimetric sensitivity is not compromised by the on-board data processing. Furthermore, we also found that the errors in the physical parameters are within the numerical accuracy of typical RTE inversions with a Milne-Eddington approximation of the atmosphere.

\textbf{Conclusion:} This paper demonstrates that the on-board data reduction of the data from \sophi{} does not compromise the accuracy of the processing. This places on-board data processing as a viable alternative for future scientific instruments that would need more telemetry than many missions are able to provide, in particular those in deep space.}




\maketitle

%%%%%%%%%%%%%%%%%
%% Sections
%
% \section{}%\label{s:?} 

\section{Introduction}
The Polarimetric and Helioseismic Imager \citep[\instr{};][]{Solanki2020PHI} is one of the instruments on-board the Solar Orbiter mission \citep[\so{};][]{Muller2020SO}. Solar Orbiter is following heliocentric orbits, that incline relative to the ecliptic plane to access higher solar latitudes. \sophi{} is a spectropolarimeter scanning the photospheric \ion{Fe}{I}\,$617.43$\,nm absorption line at two different spatial resolutions, through two telescopes: the Full Disc Telescope (FDT) and the High Resolution Telescope (HRT). The HRT is stabilised with an image stabilisation system, that corrects spacecraft jitter and follows the observed features, counteracting solar rotation. \sophi{} samples the spectral line at 6 wavelengths, recording four polarisation states at each wavelength, from which the full Stokes vector ($I$, $Q$, $U$ and $V$), describing the polarisation of the light, can be derived. To obtain a data set with reliable signal-to-noise ratio (S/N), and offer possibilities for trade-offs between S/N and acquisition time, the instrument can parametrise its acquisition scheme. The two most commonly used acquisition schemes are: (1) scanning through the absorption line while recording each of the four polarisation states five times, and accumulating four images in each state, which is completed in less than $100$\,s or (2) scanning through the spectral line and polarimetric states a single time, accumulating 16 images in each state, completed in less than $60$\,s \citep[see][]{Solanki2020PHI}. Each data set results in twenty-four images and provides information about the magnetic field vector and the line of sight velocity at an average formation height of the spectral line. We arrive at these quantities, describing the solar atmosphere, on-board the spacecraft, through a full data reduction pipeline, including the inversion of the radiative transfer equation of polarised light (RTE), assuming a Milne-Eddington approximation of the solar atmosphere. We complement the output of the inversion with the total intensity image from the continuum region next to the absorption line, as well as with metadata about all the details of the data reduction, forming the science-ready data product that is made available to scientists.

In order to facilitate on-board processing, \sophi{} has a custom designed Digital Processing Unit \citep[see][]{Solanki2020PHI} on which we implemented a data processing software system \citep[see][]{Albert2020Autonomous, lange2017board2}. There were three major drivers in the design of the data processing system: the resource limitations of the hardware, the need for autonomy of the data processing due to the long telecommand to telemetry turnaround times, and the need for the robustness of the system (i.e., to ensure complete and correct data reduction on images from different orbital positions and different solar scenes). To meet the needs with the limited resources, we had to trade off algorithm complexity and computational accuracy.

In this paper, we analyse the effect of these trade-offs on the accuracy of the data reduction pipeline. We compare processing results of a synthetic data set on a representative hardware model of \sophi{} with a reference implementation of the data reduction without trade-offs, which represents the best possible results for the data set. We use synthetic data to exclude errors from sources outside the processing pipeline; these are crucial to analyse, however, they lie outside the scope of this paper. We show that errors accumulated during the processing are negligible, and therefore we achieve the desired quality for the reduced data. We analyse the quality of the Stokes vector achieved by the on-board processing pipeline, the final accuracy of the output data, and the errors introduced during the processing.

%--------------------------------------------------------------------
\section{The on-board data processing}\label{Sec:Method}

\begin{figure}[tbp]
   \resizebox{\hsize}{!}
            {\includegraphics[width=.8\textwidth]{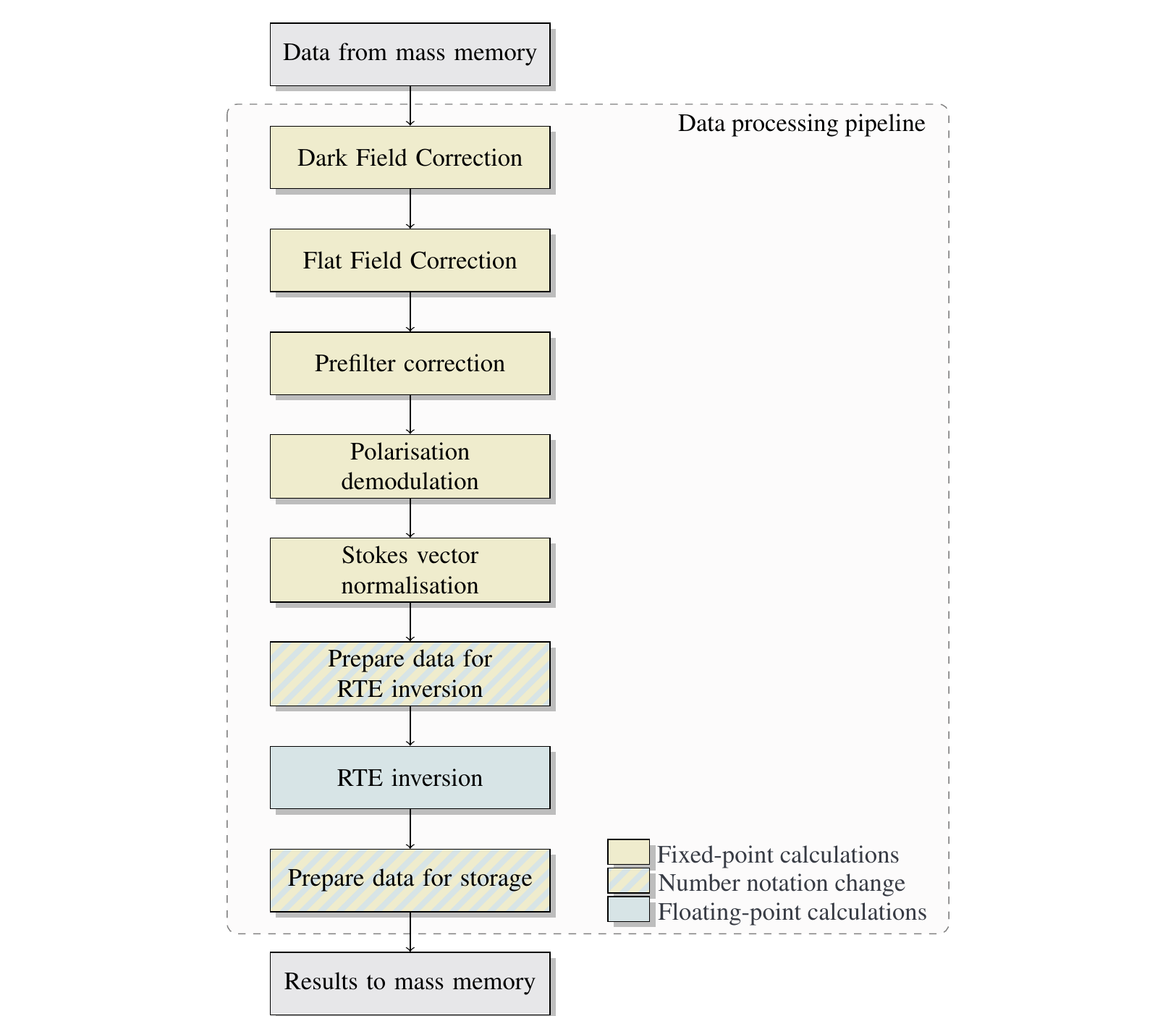}}
      \caption{The current baseline data processing of \sophi{}, executed on-board the spacecraft. To enable the calculations on the limited on-board resources, we combine fixed-point and floating-point number representation in the pipeline.}
         \label{Fig:ProcLogic}
   \end{figure}

The baseline data processing of \sophi{} consists of the standard spectropolarimetric data reduction steps (see Fig.~\ref{Fig:ProcLogic}). The processing pipeline operates on data loaded from mass memory, where we store the acquired raw data. This is also where we store the results of the pipeline, while they wait for data compression and download.
   
The pipeline starts with dark field and flat field correction. For both of these steps we determine the calibration data (i.e., the dark- and flat fields) on-board, by two separate processes, and store them in the mass memory prior to the initiation of the data reduction. Hence, the only action performed by the pipeline is their loading and their application to the data.

The following step is the prefilter correction. The exact prefilter profiles have been determined on ground at 49 different wavelengths, given in 49 different voltages of the Filtergraph \citep[see][]{Solanki2020PHI, DominguezTagle2014Filtergraph}, and uploaded to \sophi{}. The orbits of \so{} induce a continuous change in the radial velocity of our instrument with respect to the Sun, therefore we determine the voltages for data acquisition as part of instrument calibrations, on-board. This is done such that a reference wavelength $\lambda_0$, falls close to the minimum of the spectral line, and one sample falls into the nearby continuum. At the time of the data processing, we calculate the corresponding values of the prefilter profile for the data set by linear extrapolation of the measured ones, and then apply them to the data.

The polarimetric demodulation is the step that recovers the Stokes vector from the observations. For this step, the pipeline can either use a field dependent demodulation matrix, or one that is uniform across the field of view \citep[FOV; see][]{Solanki2020PHI}. Preliminary analysis performed on data retrieved up to date from \sophi{} shows that the results are more accurate with a uniform demodulation matrix. Hence, the on-board pipeline currently uses the average of the FOV dependent demodulation matrix, which was measured during the ground testing prior to launch. This is also the demodulation matrix used in this paper. Further improvement to the data demodulation is possible through polarimetric ad-hoc cross-talk correction between the different components of the Stokes vector \citep[see][]{Sanchez1992Observation,Schlichenmaier2002Spectropolarimetry}. After the demodulation and the cross-talk correction, the pipeline normalises the resulting Stokes vector, using the disc centre continuum intensity. The latter is determined on ground and uploaded to \sophi{}.

To retrieve the physical quantities, we execute the RTE inversion on the normalised Stokes vector. The RTE inversion implemented on-board uses the Milne-Eddington model atmosphere and the Levenberg-Marquardt minimisation method \citep[see][]{carrascosa2016rte}. The RTE inversion operates on a pixel basis, and needs all values of a pixel from the 24 different images. The pipeline performs this pixel sorting prior to the RTE inversion. The initial conditions for the inversion can be provided in a configuration file, or through numerical calculations, called classical estimates \citep{Semel1967Contribution, Rees1979LineFormation, Landi2004Polarisation}. After the inversion we arrive at the physical quantities: the magnetic vector, $\vec B = (\lvert\vec B\lvert, \gamma, \phi)$ where $\lvert\vec B\lvert$ stands for the field strength and $\gamma$ and $\phi$ for the inclination and azimuth of the field, respectively, and line of sight velocity, $v_{\rm LOS}$. They are then sorted back into images. Finally, the pipeline attaches the continuum intensity to them and stores them into the mass memory as the result of the data reduction.

This baseline can be further extended with other modules, such as binning and cropping, used to balance the telemetry volume with the needs of each science case. We also have the capability to further extend the pipeline in the future, e.g., with Fourier filtering to restore the data from optical effects, such as a known point spread function (PSF). 

\sophi{} combines different number representations in the data processing (see Fig.~\ref{Fig:ProcLogic}). Wherever it was possible, we opted for fixed-point representations as a method to reduce resource usage, trading off accuracy, which is one of the most important sources of accuracy loss in the pipeline. We use floating-point calculations where the accuracy of fixed-point computations did not fulfil the requirements (for instance, the inversion of the RTE). In contrast to floating-point, where number normalisation is inherent in the notation and therefore the precision is better preserved, in fixed-point representation the decimal point is always in the same place, resulting in effectively different number of bits used for the representation of different magnitudes, with $0$-padding. For instance, two irrational numbers differing only by a scale factor of $2$, both within the range of possible numbers on the allocated bits (i.e., none of them produces overflow), would have an accuracy difference of a bit in fixed point representation, while in floating point they would have the same accuracy. Maximising the accuracy of fixed point representation is possible through scaling up values, to effectively use as many bits as possible. \sophi{} uses in its data processing $24.8$\,fixed-point notation, where 24 bits are for the integer part and 8 for the decimal, and single precision, 32 bits floating point.

In order to maximise \sophi{}'s processing accuracy, we must make sure that we effectively use all the available bits in the fixed point notation by controlling the magnitude of our data in all steps of the data processing. We always scale the full data set together, to maintain the information in all dimensions of the data: spatial, spectral, and polarimetric. The individual pixel values in the images have no physical meaning throughout the pipeline, it is the normalisation of the Stokes vector by the disc centre continuum quiet Sun intensity (denoted simply as $I_c$) that creates the suitable input to the RTE inversion module.

Our starting point for scaling the data through the pipeline is in the detector. The exposure time adjusts the brightness of the solar scene such, that it is reliably represented on 12 bits read out from the detector. Then, we accumulate a number of detector readouts to increase the signal-to-noise ratio of the solar data, and pad the result with zeros after the decimal, to reach the $24.8$\,fixed point notation. Then, we calculate the largest possible integer number that we can obtain through these operations, called from here on \emph{maximum range}, and place it into the metadata of the data set for further reference. For instance, for 20 accumulations, the maximum range would be $20 \times 2^{12}$. At the start of the data processing, just after loading our raw data set from the mass memory, we scale up the data to use all the available bits. This is achieved by multiplying with a scale factor calculated as the ratio of $2^{23}$ (the largest possible integer in two's complement on $24.8$\,fixed-point notation) and the current maximum range. In each operation that follows, we aim to preserve this largest possible maximum range in the result, by considering the magnitude of the operands and that of the result. For simplicity, we only keep track of the maximum range. However, in few cases the minimum range of the absolute value of the data is also relevant, for instance,  when we perform divisions like that necessary to correct the flat field or the prefilter. Here, the smallest value in the divisor determines the maximum range of the result. Since these are not tracked, we make assumptions about the divisor, with the consequence that any pixels with smaller values will create an overflow and the resulting pixel will become not-a-number (NaN).

We scale all calibration data to no higher value than to represent the precision with which they are determined (e.g., in the case of the flat fields we only use $3.8$\,bits). In those cases where the accuracy of the calibration data is high, we do a trade-off between the accuracy of the data and that of the calibration data. We decide all trade-offs on a case by case basis, based on simulations to verify which scaling gives the best precision results.

\section{Test setup}\label{Sec:TestSetup}
To compare the accuracy of the on-board processing pipeline to what could be achieved on-ground, and isolate it from other sources of errors (e.g., solar evolution, and the accuracy of the calibration), we have chosen to process a synthetic data set. We process this data with the pipeline described in Sect.~\ref{Sec:Method} on the Qualification Model (QM) of \sophi{}, which is fully representative in terms of the Data Processing Unit of the Flight Model. We compare the QM results to a reference pipeline run in a PC, in double precision floating point, representing the best possible processing accuracy. The one exception is the RTE inversion, which is performed on the QM in both cases. In the case of the reference pipeline, we upload the RTE input data in floating point and download directly the floating point results that it produces. The data processing pipeline uses the same calibration data that we apply in the data preparation, therefore we have no inaccuracy originating from the determination method and processing of the calibration data. This means that all errors presented in this work are inaccuracies originating in the data reduction pipeline.

   \begin{figure}[tbp]
   \centering
   \includegraphics[width=0.97\textwidth]{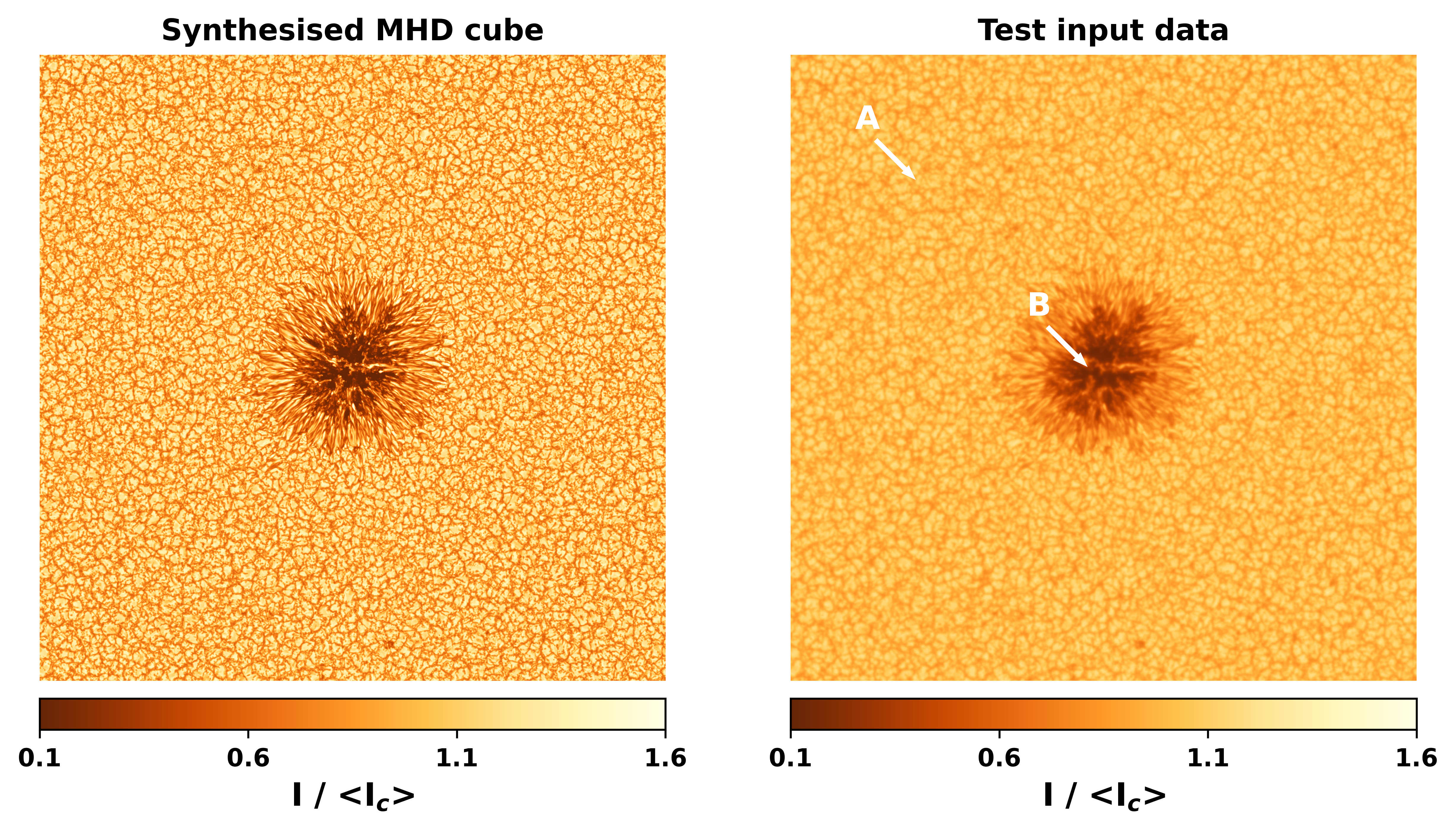}
      \caption{The data set used in the tests contains an MHD simulated sunspot \citep{Rempel2015Numerical}. We synthesised the spectral line and its nearby continuum from the MHD cube (left) and degraded, as described in the main text, to what we would expect from \sophi{} (right) shown here at the continuum intensity sample wavelength, $\lambda_0 - 300$\,m\AA. The most obvious result of the degradation is the loss of RMS contrast. The size of the data is $1024 \times 1024$ pixels, each pixel corresponding to $48$\,km on the Sun. The spatial sampling is larger  than what \sophi{} achieves at closest approach. However, it is preserved to provide more pixels for statistical analysis. The arrows indicate pixels for which the Stokes profiles are plotted in Figs.~\ref{Fig:InputData_a} and \ref{Fig:InputData_b}.}
         \label{Fig:InputData_location}
   \end{figure}

The test data set is a magnetohydrodynamic (MHD) simulation of a sunspot \citep{Rempel2015Numerical}. From this MHD cube we synthesized the \ion{Fe}{I}\,$6173$\,\AA\ spectral line profiles, applying a wavelength sampling of $14$\,m\AA, with the SPINOR code \citep{Frutiger2000Properties}. SPINOR relies on the STROPO routines to solve the RTE \citep{Solanki1987PhDT}. The simulation is $1024 \times 1024$ pixels, with a pixel size of $48$\,km. The pixel size of the HRT telescope of \sophi{}, at closest approach, corresponds to $101$\,km, however the dataset is not resampled, in order to provide more pixels in the umbra and penumbra for statistical analysis.

We degrade the synthesized spectral line profiles in several steps, starting by convolving the wavelength dimension of the synthesised data with the transmission profile of the \sophi{} Filtergraph \citep[see][]{Solanki2020PHI, DominguezTagle2014Filtergraph}:
\begin{equation}
    S^{conv}_{p}(\lambda,x,y) = S^{synth}_{p}(\lambda,x,y) \ast F(\lambda),
\end{equation}
where $\ast$ denotes convolution, the index $p$ runs over the four polarimetric modulation states, $\lambda$ denotes the wavelength, $x$ and $y$ are the spatial image coordinates in pixels. $S^{synth}$ are the Stokes profiles synthesized from the MHD cube, $F$ denotes the filter profile, and $S^{conv}$ is the Stokes vector from the synthesis, convolved with the spectral profile.
We then select the samples for \sophi{} from the resulting spectral profiles. The samples are defined relative to a reference wavelength ($\lambda_0$, which we chose for this test to be $6173.371$\,\AA). $\lambda_0$ is placed in the vicinity of the absorption line minima:
\begin{equation}
\begin{split}
    &S^{samp}_{p}(\lambda,x,y) = S^{conv}_{p}(\lambda_{s},x,y),
\end{split}
\end{equation}
where $\lambda_{s}$ denotes the sample wavelengths in reference to $\lambda_{0}$: \[\lambda_{s} = [\lambda_0 -300\,m{\rm\AA{}},\lambda_0 -140\,m{\rm\AA{}},\lambda_0 -70\,m{\rm\AA{}},\lambda_0,\lambda_0 +70\,m{\rm\AA{}},\lambda_0 +140\,m{\rm\AA{}}],\] and $\vec S^{samp}$ is the Stokes vector sampled in wavelength.

After spectral sampling, we convolve each individual image of the data set (the $24$ images, six wavelength samples and four modulation states) with the spatial PSF in the shape of a Lorentzian function, configured with the theoretical parameters of the HRT telescope, adjusted to the plate scale of the simulation:
\begin{equation}
     S_{p}(\lambda,x,y) = S^{samp}_{p}(\lambda,x,y) \ast A(x,y),
\end{equation}\label{Eq:stokes}
where $A$ represents the PSF of the \sophi{} HRT.

The convolution with the PSF reduces the continuum quiet Sun root-mean-square (RMS) contrast of the synthetic data from $22.83\%$ to $7.9\%$. The next step is the polarimetric modulation of the synthetic data:
\begin{equation}
      I^{mod}_{m}(\lambda,x,y) = k\sum_{p=1}^{4} M_{mp}(\lambda) S_{p}(\lambda,x,y),
\end{equation}
where index m denotes modulation states, $I^{mod}_m$ is the modulated data set, $M_{mp}$ is the polarimetric modulation matrix, and $S_{p}$ is the Stokes vector from Equation~(\ref{Eq:stokes}). The constant $k$ adjusts the quiet Sun continuum mean intensity ($I_c$) of the data to what is representative of the \sophi{} HRT, converts the number of photons collected by the detector to digital numbers, read out from the electronics. This constant, furthermore, accounts for the frame accumulations that we perform in order to increase the signal-to-noise ratio of the data, which we adjust to $20$ frames for the tests. Due to the higher RMS contrast in the test data, and therefore higher dynamic range, we adjust the $I_c$ slightly below the level observed in the \sophi{} HRT data. The ratio of $I_c$ for the observed to the test data is $1.08$. It is important to remark that this difference in $I_c$ creates a slightly worse case from the point of view of accuracy: as the values in the test data are somewhat lower, they are represented on fewer bits, when compared to \sophi{} observations. We remark, that the order in which we apply the instrumental degradation to the synthetic data has been established for the sake of convenience and to be able to correct the data in the same order with the pipeline. For instance, we have applied the polarimetric modulation here, even though the modulation package in the instrument is right after the entrance window following the beam direction. Since only the order of linear operations have been exchanged (only the dark field application is non-linear, which we apply according to the optical path), this does not affect the resulting input data.

Once we have modulated the input Stokes vector and converted the data to digital numbers, we apply the prefilter profiles, dark and flat fields to the synthetic data:
\begin{equation}
      I^{obs}_{m}(\lambda,x,y) = I^{mod}_{m}(\lambda,x,y) I^\text{p}(\lambda,x,y) I^\text{f}_{m}(\lambda,x,y) + I^\text{d}(x,y),
\end{equation}
where $I_m^{obs}$ is the data set produced to match \sophi{} observations, i.e., the input to the processing pipeline, $I^\text{p}$ is the prefilter profile which is different for each wavelength and pixel, $I^\text{f}_m$ are the flat fields, depending both on wavelength and polarisation states $m$, and ${I}^\text{d}$ is the dark field of the sensor, the same for all wavelengths and polarisation states. 

It is worth noting that the addition of the flat field and dark field to the data further reduces its RMS contrast to $6.4\%$. During the first months of \sophi{} operations, we have observed the RMS contrast in \sophi{} HRT observations of the quiet Sun to be around $4.5\%$ to $5\%$, which is expected to change further as \sophi{} changes its distance to the Sun. The smaller the dynamic range in the data (i.e., lower the RMS), the closer we could take all values of it to the maximum possible range, hence achieving a better overall computational accuracy. For this, we could do an adjustment in the exposure time, or in the maximum range assumptions at data acquisition. This is, however, not done in the case of our test data. Therefore, the results are representative of what we would achieve on \sophi{} data. Also note, that the dark- and flat fields that we apply here are calculated on ground from \sophi{} data, and calculated on-board \sophi{}, respectively. These data we downloaded during the commissioning phase of the mission. We produced this early calibration data with methods that we will improve further, however these are representative of the flat- and dark fields that we expect after fine-tuning. Furthermore, it is important, that in order to not introduce further uncertainty into the process, and be able to assess the effects of the numerical errors on the RTE inversion, we omit several effects that would appear in real data. We do not introduce noise to the data in the course of its degradation. Likewise, we use the data as instantaneous snapshots of the solar scene without considering the evolution of the solar scene, rotation of the Sun or spacecraft jitter.

   \begin{figure}[tbp]
   \resizebox{\hsize}{!}
            {\includegraphics[width=\hsize]{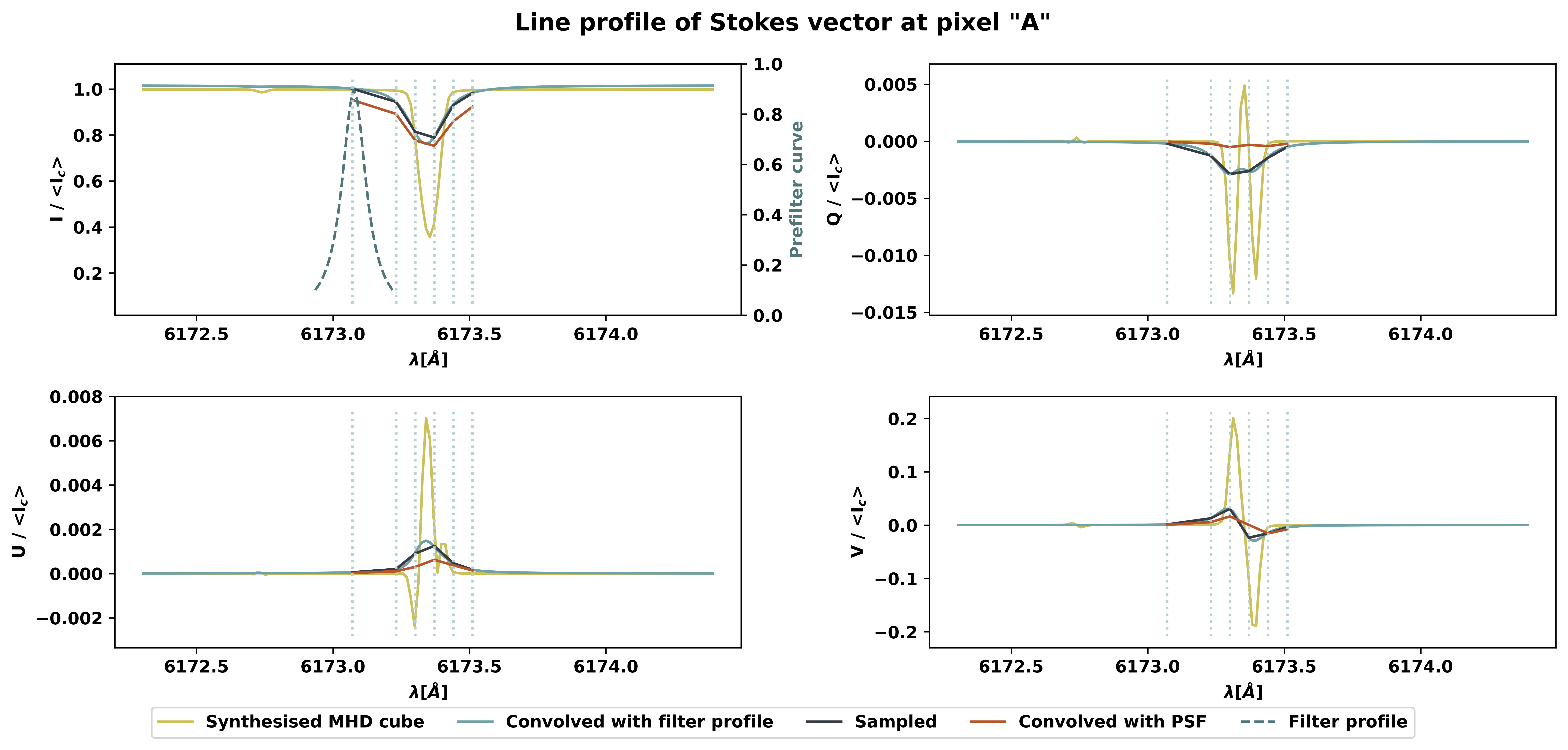}}
      \caption{The spectral line profile of the input data set shows the degradation of the synthesised MHD data. We first convolve the synthesised profile with the filter profiles of \sophi{} (shown in the top left panel), then select the correct wavelength samples, followed by the convolution of the resulting images with the theoretical PSF of the instrument. This is a bright pixel from the quiet Sun. The convolution with the filter profiles significantly reduces the spectral line complexity.
      We indicate the location of the plotted pixel in Fig.~\ref{Fig:InputData_location}.}
         \label{Fig:InputData_a}
   \end{figure}

 \begin{figure}[tbp]
   \resizebox{\hsize}{!}
            {\includegraphics[width=\hsize]{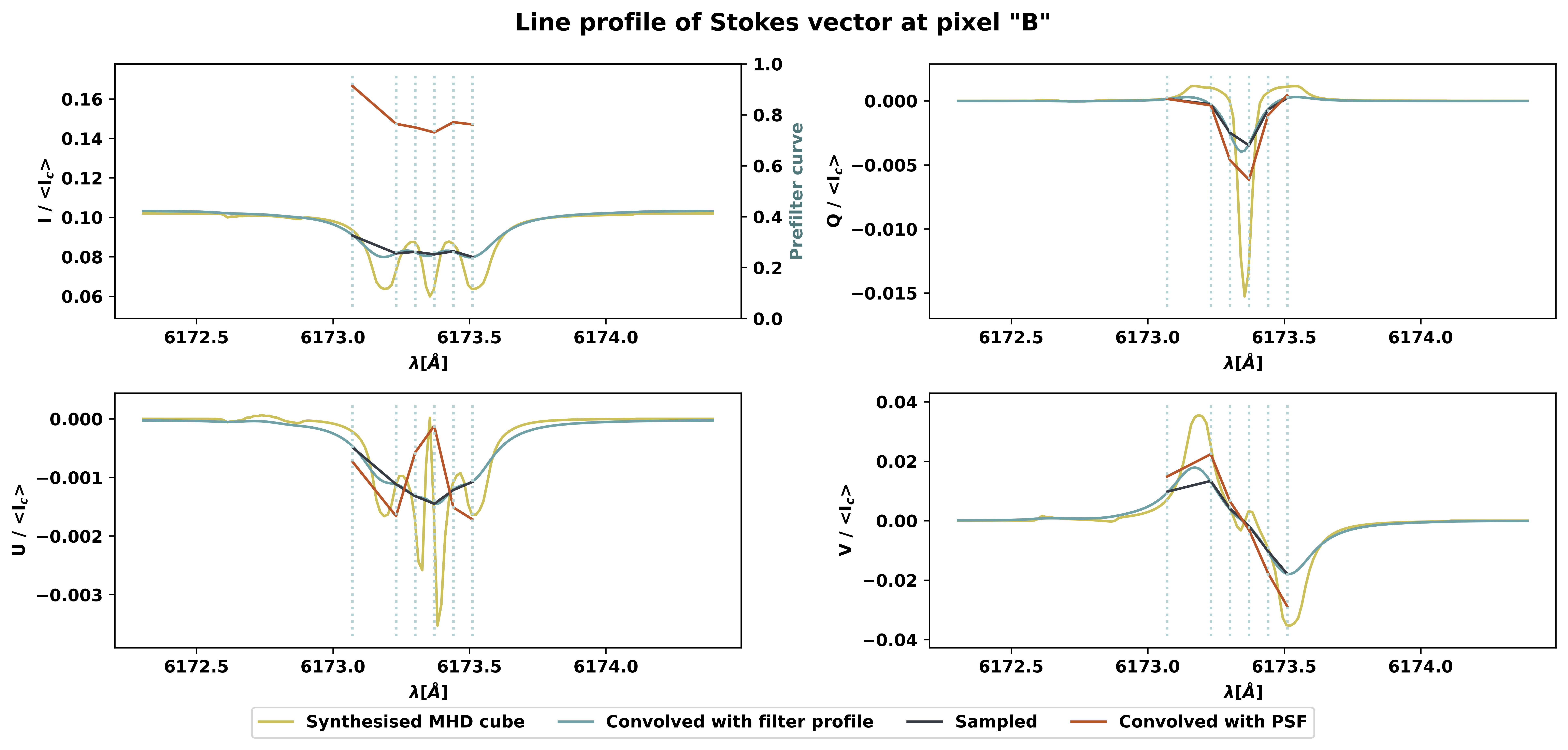}}
      \caption{
       Same as Fig.~\ref{Fig:InputData_a}, for a pixel from the umbra. This dark pixel is surrounded by bright structures, therefore after applying the PSF, there is a large change in the intensity level at continuum from neighbouring pixel contributions. There is also a significant change in the $U$ profile. This is an extreme case, with neighbouring pixels being significantly different from the one selected. Here, the PSF convolution has a much stronger effect on the final profiles, when compared to Fig.~\ref{Fig:InputData_a}. We indicate the location of the plotted pixel in Fig.~\ref{Fig:InputData_location}.}
         \label{Fig:InputData_b}
   \end{figure}
   
We show the effects of the data degradation at the continuum wavelength in Fig.~\ref{Fig:InputData_location}. The most obvious effect is the reduction of the image RMS contrast. In the line profiles of the data (see Figs.~\ref{Fig:InputData_a}, and \ref{Fig:InputData_b}) we can see that the complex profiles from the MHD simulations smooth out significantly as a result of the convolution with the transmission profiles of the \sophi{} Filtergraph. The same operation also lowers the amplitude of the polarisation signals. The sampling of the data further removes details of the spectral shape by reducing the available information. This effect is especially strong in the sunspot profiles, due to the complex shapes. The spatial PSF convolution strongly changes the Stokes I intensity, especially in the umbral profile, as usually stray light does in real observations. We do not expect the RTE inversion to perfectly reconstruct the resulting profiles, producing especially large differences in the umbra, as it does not account for the stray light. The spatial PSF can lower the amplitude of the polarisation signals further (e.g., in quiet Sun areas) although it is not always the case (as shown here in the umbral profiles) since the final effect on each pixel depends on the surrounding signals. Note, that the final, degraded profiles appear different with respect to the synthesised data. This is mainly an effect of the wavelength sampling. While we sample the absorption line with symmetric offsets, the reference wavelength does not necessarily coincide with the centre of the line, causing a shift in the sampling, and introducing an apparent asymmetry even in the case of symmetric profiles. This, however, does not affect the performance of the RTE inversion. The relatively few samples (only five) also contribute to the strong difference.

We analyse how the data pipeline changes the accuracy of the data throughout each step of the pipeline, grouping them into three categories: the input and the first steps of pre-processing, the polarimetric errors, and the physical parameters. In the first category, we start by analysing the input errors introduced by the data quantisation to fixed point notation. Next, we look at the early pre-processing errors, which are introduced by the first three pipeline steps: dark field, flat field and prefilter correction. The next category comprises the polarimetric sensitivity, where we evaluate the errors in the determination of the Stokes vector during the demodulation process and the cross-talk correction. Achieving a good polarimetric sensitivity of the Stokes vector is the most important requirement of \sophi{}. Finally, in the third category we give a glimpse into the resulting physical parameters, in which we analyse the results of the inversion. These results correspond to the science ready data obtained directly on-board.

 \begin{figure}[tbp]
   \centering
   \includegraphics[width=\hsize]{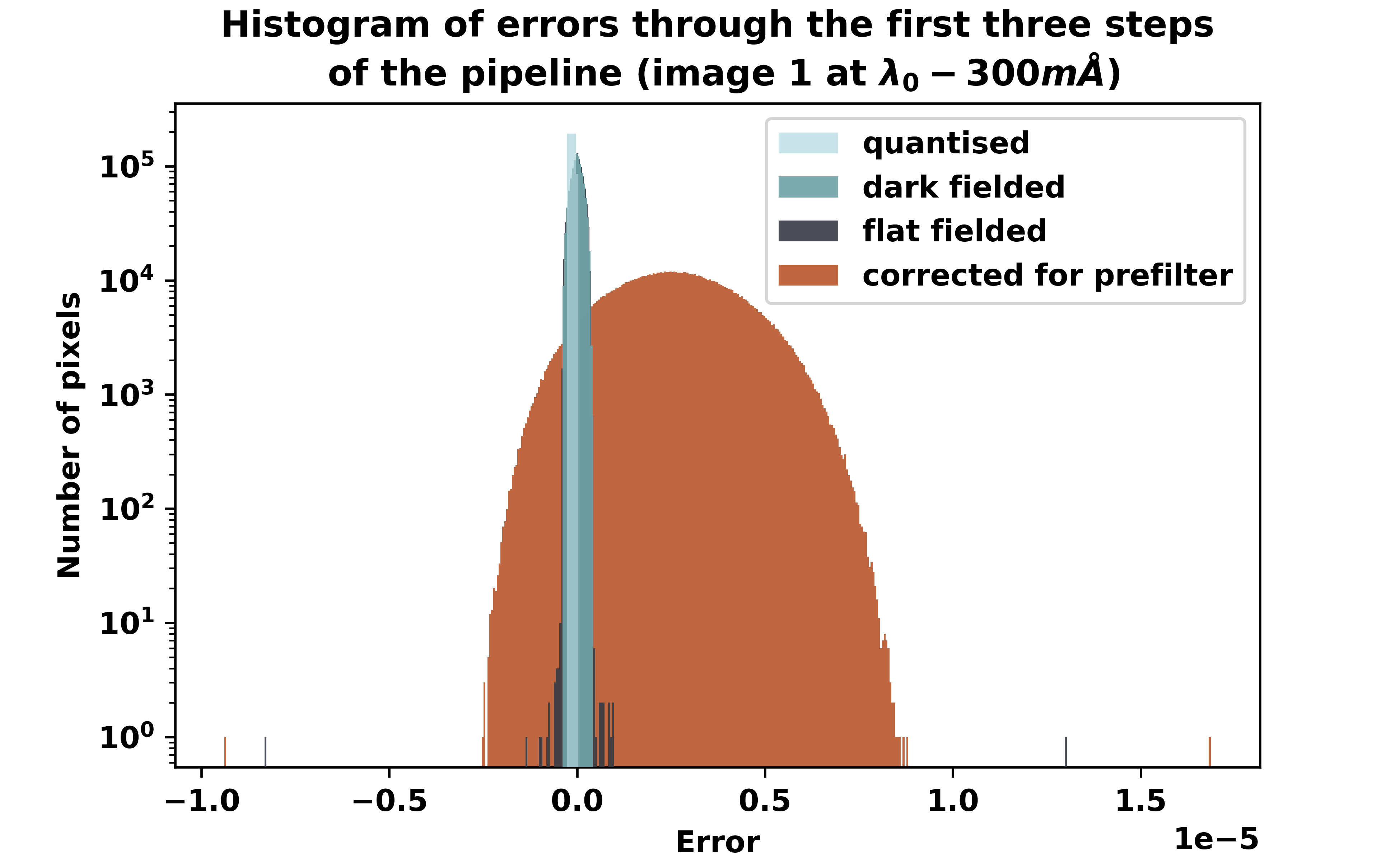}
      \caption{The histogram of errors after data quantisation and the first three steps of the pipeline shows the accuracy achieved after each step, normalised to the mean of the data. The quantisation and dark field subtraction introduces small errors. The errors introduced by the flat field division are very similar to that of the dark field correction, however a few outliers show up from dust grains in the FOV. The prefilter correction produces a residual of the image with very low intensity due to the inaccuracy at the extrapolation of the prefilter profile to the correct voltage.}
         \label{Fig:Histogram_Total}
   \end{figure}

\section{Analysis}

For the analysis, we separate the data into three zones, based on the Stokes vector signal levels: the umbra, the penumbra and the quiet Sun. The regions are defined on the conservative side, with preference on excluding pixels from them, rather than including pixels that do not clearly belong. We consider $>10000$, $>81000$, and $>94000$ pixels in the three regions, which corresponds to $1\%$, $7.8\%$ and $90.5\%$ of the full field of view, respectively. We follow this definition in the rest of the paper, and mark these regions on the figures.

\subsection{Input and first steps of pre-processing}

The first source of error is the quantisation error of the input data to the pipeline. After calculating them in double precision floating-point, according to the description in Sect.~\ref{Sec:TestSetup}, we transform the data set to the fixed-point representation as the raw data would be stored: detector read-out on $12$ bits, accumulated 20 times, and padded with 0-s for the decimals. This corresponds to a maximum range of $20 \times 2^{12}$. The RMS of the error resulting from the quantisation, normalised to the image mean intensity, is between $1.55 \times 10^{-7}$ and $3.5 \times 10^{-7}$ across the FOV of the 24 images (the six different wavelengths and four polarisation states). This is consistent with the $1 \div 2^8$ precision of the decimal in the $24.8$ fixed point representation. The profile of the histogram is flat, as expected for quantisation errors (see Fig.~\ref{Fig:Histogram_Total}), with a systematic bias towards smaller numbers in the quantised data, due to using bit truncation, rather than rounding.

The first steps of pre-processing are: the subtraction of the dark field from the data, the division of the data by the flat field, and the division of the data by the extrapolated prefilter profiles. After the loading of the data, the pipeline scales it up by $2^{23} \div (20 \times 2^{12}) = 102$, which is also followed by the scaling of the dark field. The errors after this step originate from data quantisation both in the input (as described before) and the dark field. The subtraction operation itself does not produce any errors, inherently. After this step, we have an error RMS ranging between $1.55 \times 10^{-7}$ to $7.7 \times 10^{-7}$, normalised to the image mean in the 24 images of the data set. This step slightly changes the profile of the error histogram, removing the bias caused previously by bit truncation (see Fig.~\ref{Fig:Histogram_Total}). 

The following step, the division by the flat field, divides the data with a max range $2^{23}$, i.e., effectively using $24.8$ bits, by data with max range $2^3$, i.e., effectively using $3.8$ bits. The flat field has been normalised to its mean intensity, scaled by $2^3$, and we assume its minimum range to be $2^2$. Any number below this may cause an overflow, however there is still some room for smaller values, as the data does not fill up the full detector well at acquisition. After the division, we readjust the magnitude of the result to $2^{23}$ by multiplying it with $2^2$. The errors in this step originate from the errors on the input data (as shown in previous steps), the representation error of the divisor, and the representation errors of the result. In the histogram, a few pixels with larger errors appear due to dust grains in the FOV (which translates to very small numbers in the divisor). However, these errors are only in a handful of pixels, not contributing significantly to the RMS calculated over the full FOV, which ranges between $1.56 \times 10^{-7}$ and $7.7 \times 10^{-7}$, normalised to the image mean intensities. 

The next step, the prefilter correction, starts with the interpolation of the prefilter. The pipeline performs this on data scaled to a maximum range of $2^{23}$, then scales it down to a maximum range of $2^{10}$. Furthermore, we assume a $2^9$ minimum range. The interpolation of the data creates an error in the divisor, compared to what we obtain on ground. Then, through the division, the error histogram widens: when subtracting the results, a very small amplitude residual of the data remains. The histogram profile seen in Fig.~\ref{Fig:Histogram_Total} is the histogram of the test data. After the operation, we scale the result of the division back to $2^{23}$ by multiplying it with $2^9$. The RMS of the error across the FOV ranges between $3.1 \times 10^{-6}$ and $3.75 \times 10^{-6}$, normalised to the mean intensity of the images.

   \begin{figure}[tbp]
   \centering
   \includegraphics[width=.8\textwidth]{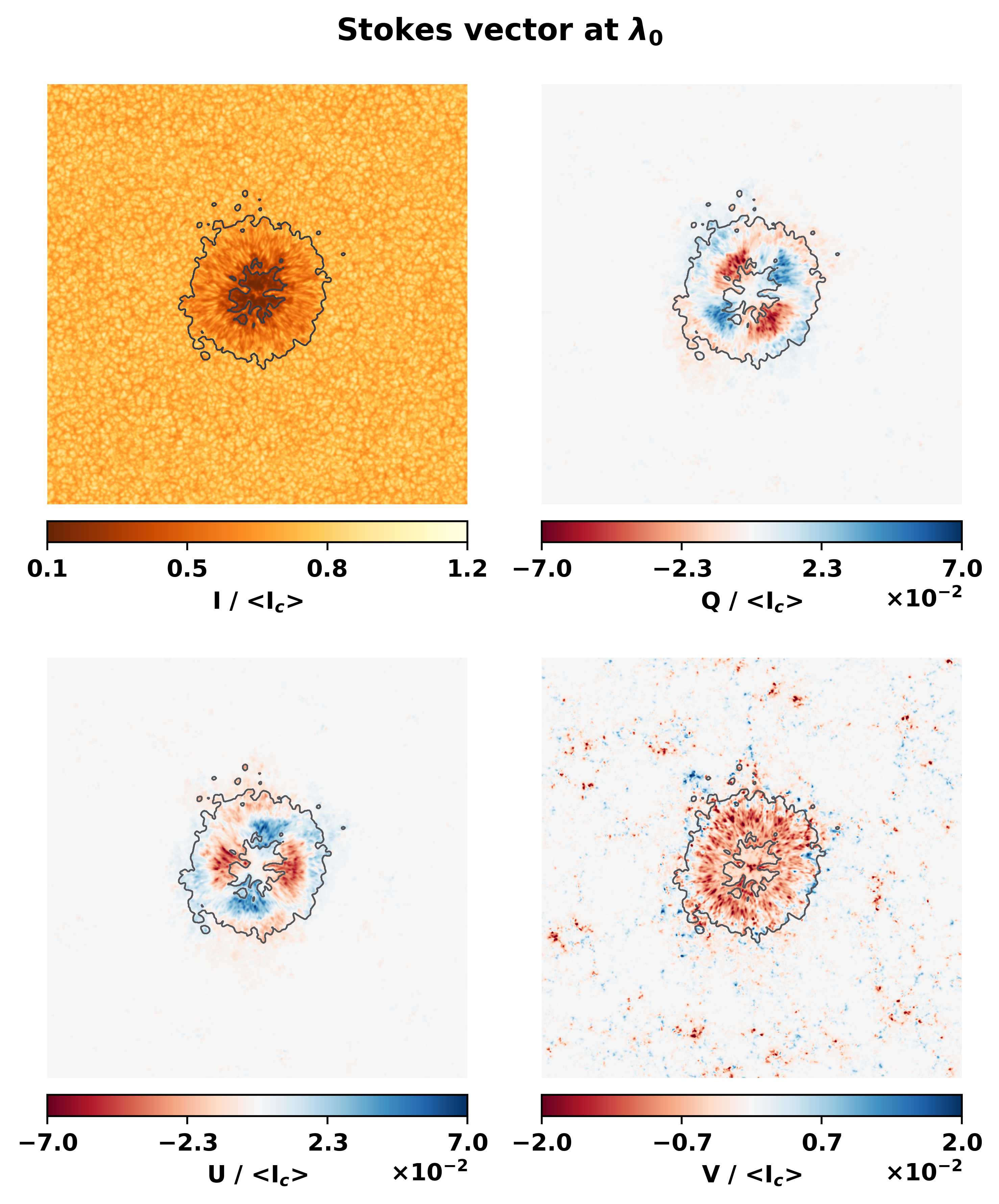}
      \caption{The normalised Stokes vector, $\vec S/I_c$ = ($I$, $Q$, $U$, $V$)$/I_c$, is shown here at the reference wavelength, $\lambda_0$ (6173.371\,\AA), close to the minimum of the absorption line. As expected, Stokes $Q/I_c$ and $U/I_c$ shows the strongest signals in the penumbra. This is also true for $V/I_c$, which is due to the large splitting of the line in the umbra, causing a poor sampling by \sophi{}. (see Fig.~\ref{Fig:InputData_a}).}
         \label{Fig:DataThroughPipeline_Stokes}
   \end{figure}

\subsection{Polarimetric errors}
   \begin{figure}[tbp]
   \centering
    \begin{minipage}[c]{\hsize}
    \centering
    
    \includegraphics[width=.65\hsize, angle=-90]{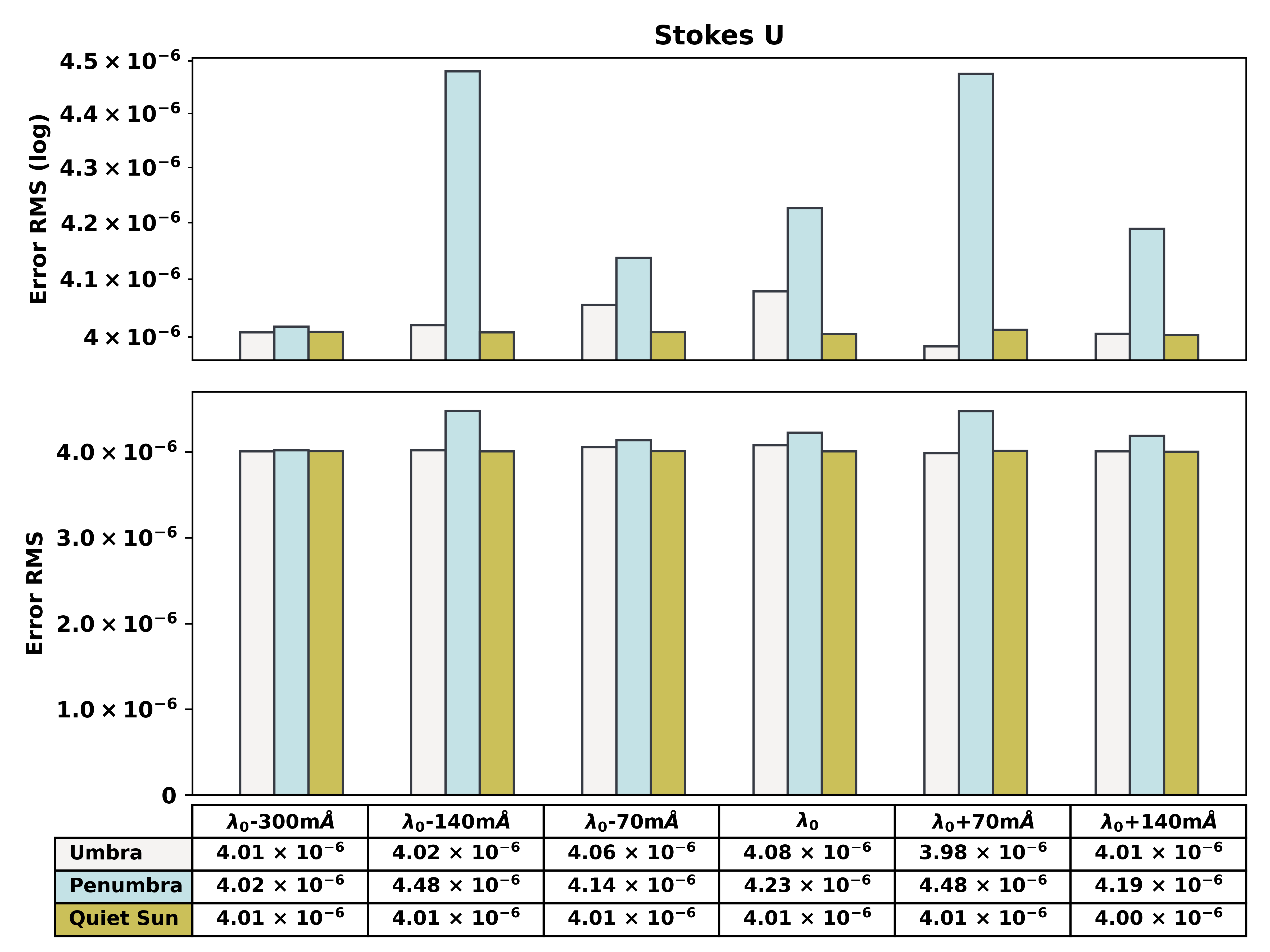}
    \includegraphics[width=.65\hsize, angle=-90]{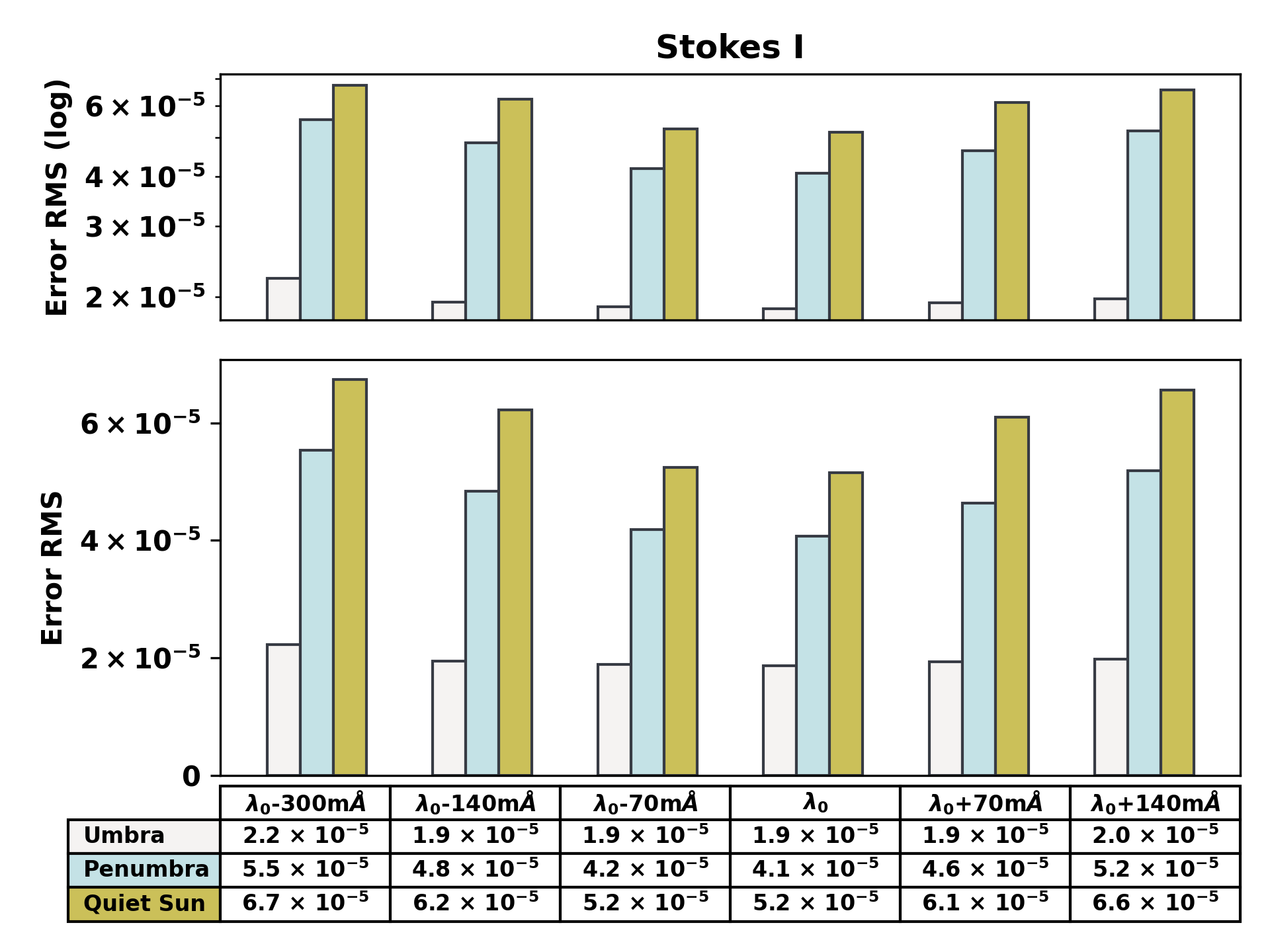}\\
    \includegraphics[width=.65\hsize, 
    angle=-90]{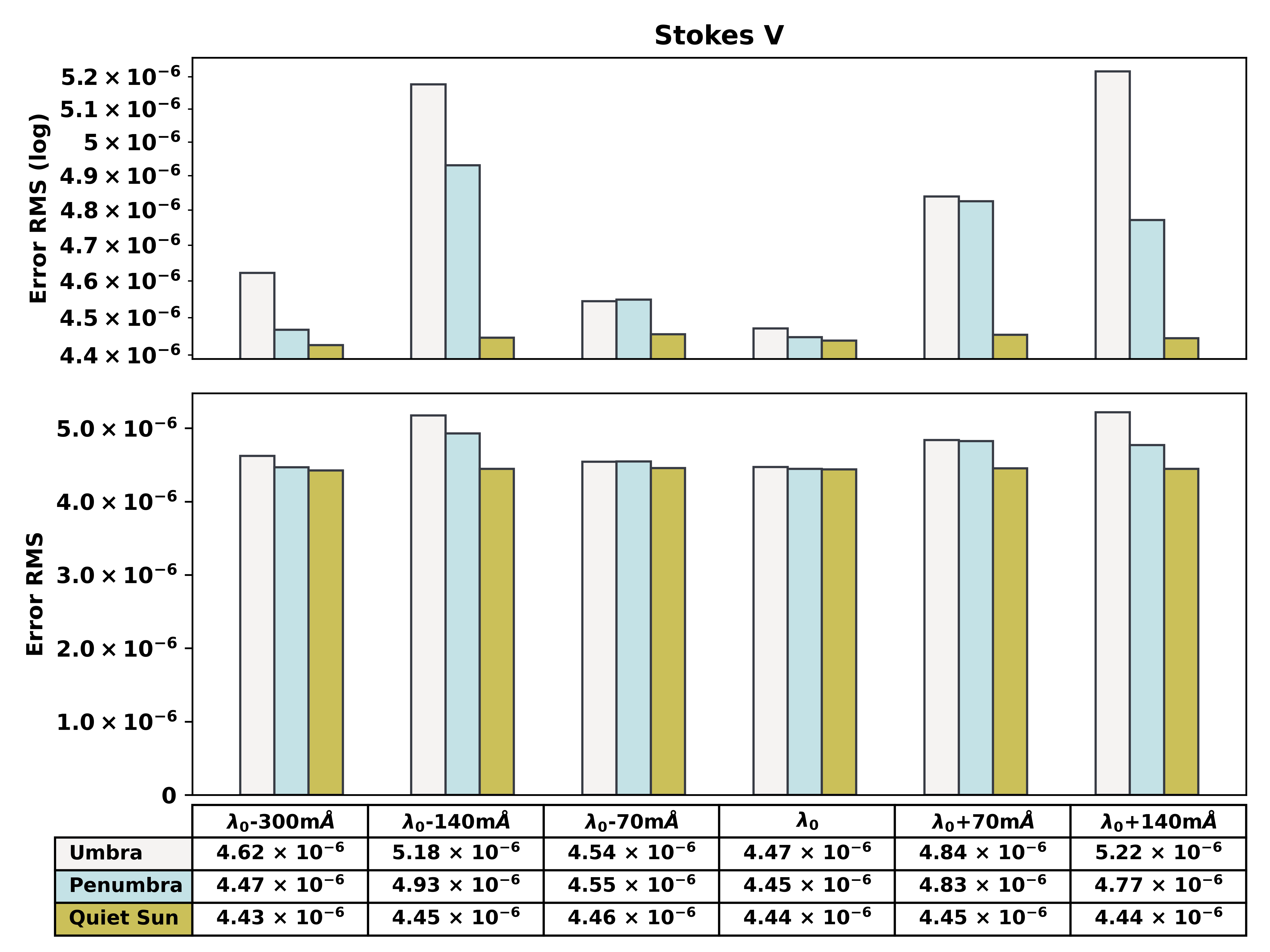}
    \includegraphics[width=.65\hsize, angle=-90]{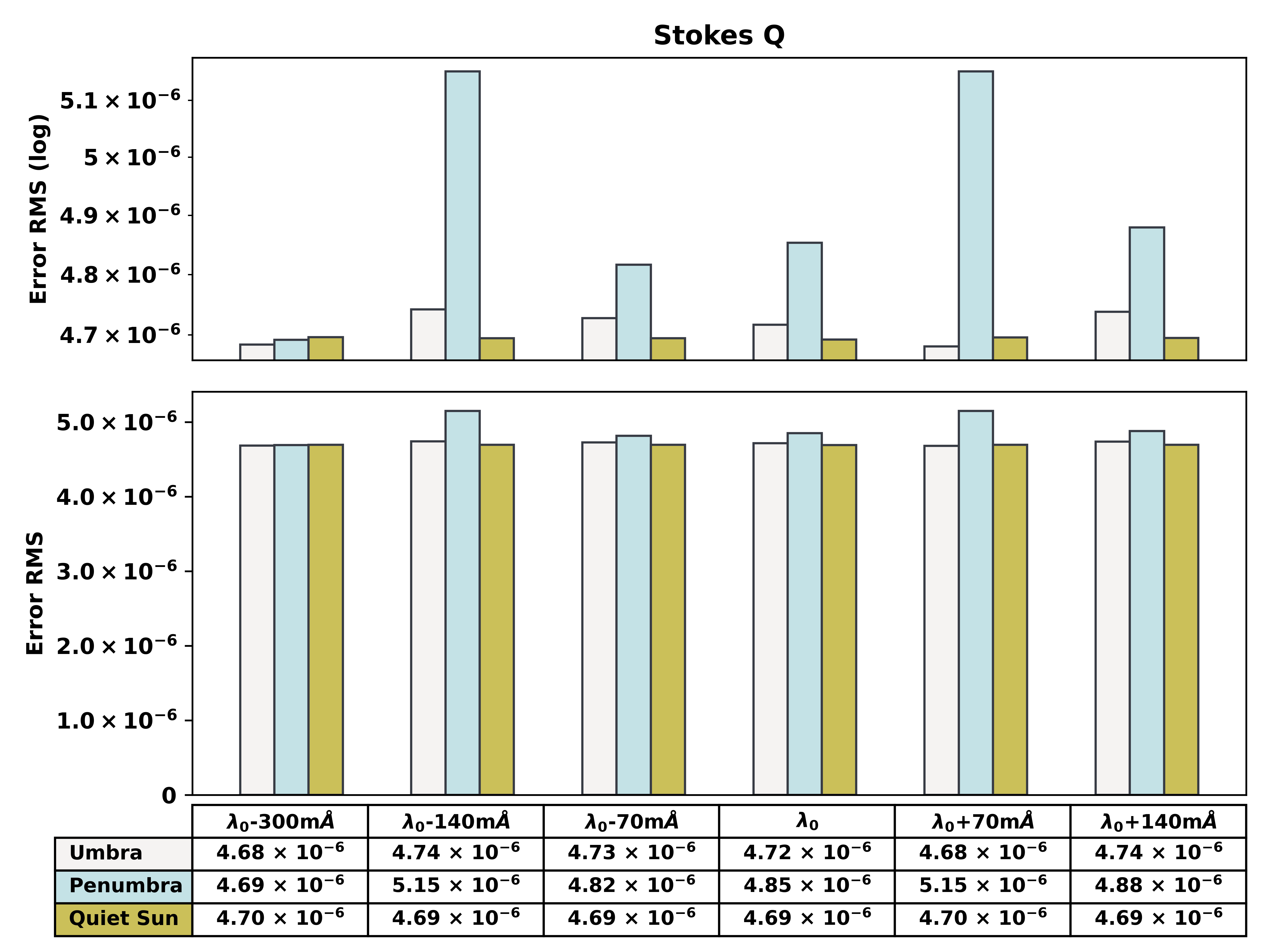}
    \end{minipage}

     \caption{The RMS of the errors in the Stokes parameters varies between $4.0 \times 10^{-6}$ and $6.6 \times 10^{-5}$, which leaves a good margin to meet the $10^{-3}$ polarimetric sensitivity requirement of \sophi{}. The errors vary by $6.5 \times 10^{-5}$ across the Stokes parameters, wavelength and solar regions, as shown in the table on the bottom. Their relation in linear scale (bottom bars) shows that Stokes $Q$, $U$, and $V$ are very close to each other, therefore to better see their differences, we use a logarithmic scale (top).}
         \label{Fig:BarPlotStokes}
    \end{figure}
    
The Stokes vector is the output of the polarimetric demodulation of the data. The pipeline is able to do further adjustments with ad-hoc polarimetric cross-talk correction methods. However, for this study, we will limit ourselves to errors due to the demodulation process. We scale the demodulation matrix to the maximum range $2^9$, for which we have to account in the input data. In order to avoid overflow, we divide the input data by $2^9 \times 4$, where the number $4$ accounts for the addition of the rows in the $4 \times 4$ matrix multiplication. The result of the operation is then on a maximum range of $2^{23}$. Since the values of the Stokes vector at this point do not have physical meaning, we normalised them to the mean of the quiet Sun intensity, $I_c$, before showing them in Fig.~\ref{Fig:DataThroughPipeline_Stokes}. The normalisation here is performed in double precision, to show the results without the error introduced by this step, when performed on-board.

The polarimetric sensitivity requirement set for \sophi{} is $10^{-3}$, which is met during the processing: the errors accumulated by the end of this step have an RMS across the FOV between $3 \times 10^{-6}$ and $4.7 \times 10^{-6}$ (see Fig.~\ref{Fig:BarPlotStokes}). This leaves a generous margin to other sources of error, and does not compromise the required polarimetric precision.

The errors across the different Stokes parameters, the different wavelengths and the different regions of the data (umbra, penumbra, and quiet Sun) differ by a maximum of $1.7 \times 10^{-6}$, which is considered negligible. The variation of error is visible for $I$ on a linear scale. However, for $Q$, $U$ and $V$ we need a logarithmic scale to illustrate the differences. The errors in the result depend on four factors: the errors that were accumulated prior to this step, the magnitude of the input data, the magnitude of the output data and the terms of the demodulation matrix. These first and second sources oppose each-other: the so far accumulated error correlates linearly with the intensity (due to the image residual after prefilter correction), while the input data is better represented where the values are larger, therefore it has an inverse correlation. The demodulation matrix has only positive terms in the row producing Stokes $I$, however it has negative terms for $Q$, $U$, and $V$, producing cancellation effects.

\begin{figure}[htbp]
   \centering
   \includegraphics[width=0.8\textwidth]{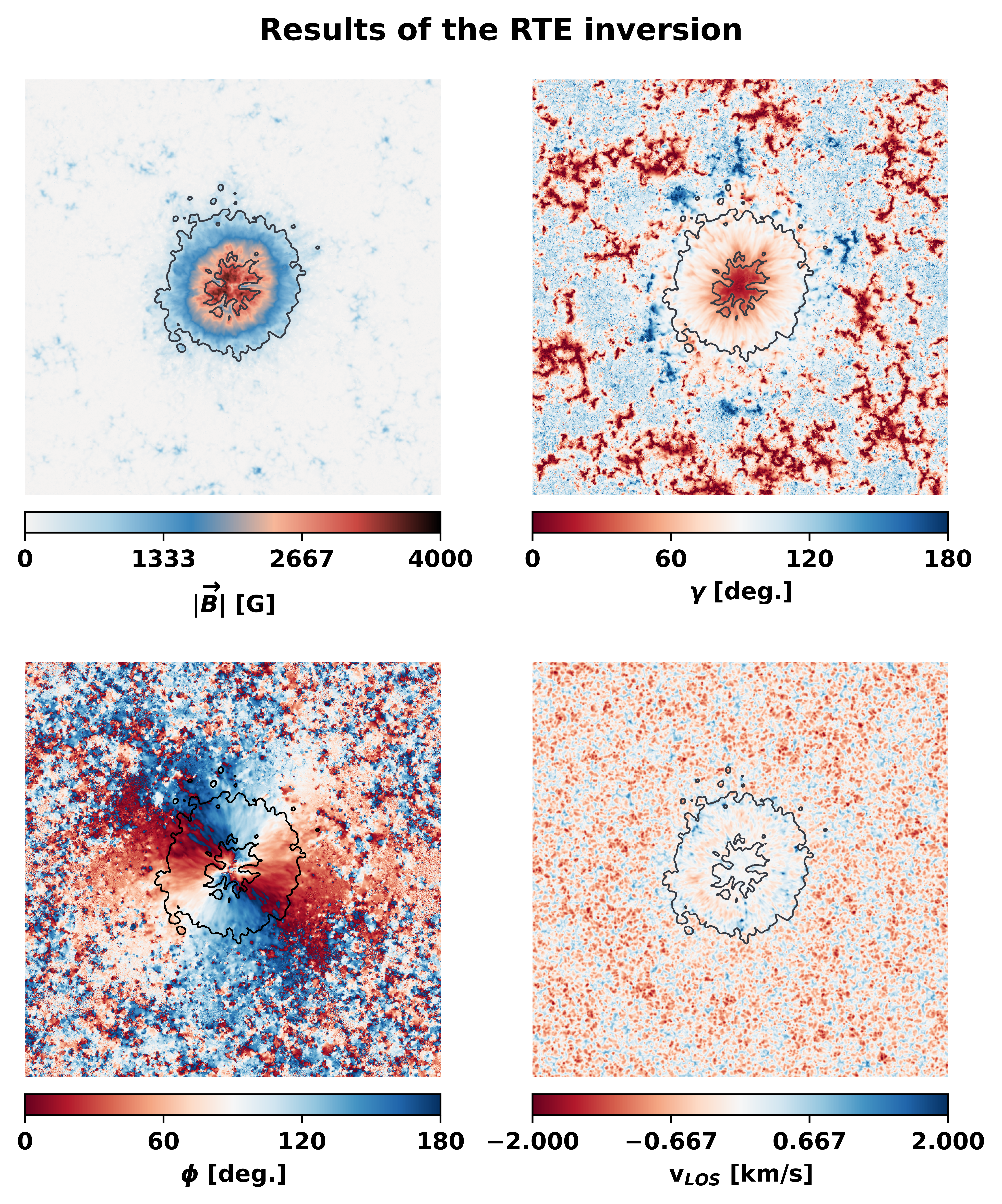}
      \caption{The results of the RTE inversion, shown here, together with the continuum intensity image, form the science ready data that is transferred to ground. The results are consistent with what is expected from such a data set: $\lvert\vec B\lvert$, $\gamma$ and $v_{\rm LOS}$ shows smooth transitions between the structures, with magnetic fields up to $4000$\,G. In $v_{\rm LOS}$ we can see the up and down flows of the solar granulation, as well as of the Evershed flow. $\phi$ is dominated by noise in the quiet Sun, however it does show the fan-like structure around the penumbra, as expected.}
         \label{Fig:Results}
   \end{figure}

The trend in errors in Stokes $I$ approximately follows the intensity of the output, with only a slight deviation from this trend along the spectrum (see Fig.~\ref{Fig:BarPlotStokes}). This is a case where the intensity in the final results dominates the magnitude of the errors. In Stokes $Q$, $U$ and $V$ the intensity of the result does not overpower the trend any more, and there is a cancellation effect of previous errors due to negative terms in the demodulation matrix. The result of all these values is a trend in errors that is stochastic.

\subsection{Physical parameters}

We reach the final physical quantities through the RTE inverter \citep{carrascosa2016rte}. The inversion module can be configured in five different modes, depending on the desired outputs, and on the initial model (apart from special debugging modes). The first three can provide all atmospheric parameters of the Milne-Eddinton model (line-to-continuum absorption coefficient ratio, Doppler width, damping coefficient, source function — its slope and its value at the top of the atmosphere, magnetic field vector $\vec B = (\lvert\vec B\lvert, \gamma, \phi)$, and LOS velocity, $v_{\rm LOS}$). These modes are: inversion starting with an initial model, called classical estimates, calculated with analytical formulae (centre of gravity technique, see \cite{Semel1967Contribution}, and the weak-field approximation, see \cite{Landi2004Polarisation}), inversion with a configurable initial model, and the classical estimates without being followed by an inversion. Aside from these modes, we have two others which only return line of sight (LOS) parameters: the longitudinal mode where we only obtain LOS velocity and LOS magnetic field, and the no polarisation modulation mode where we only obtain LOS velocity. We select the modes based on the required science return and available telemetry. 

\begin{figure}[tbp]
   \centering
   \includegraphics[width=.8\hsize]{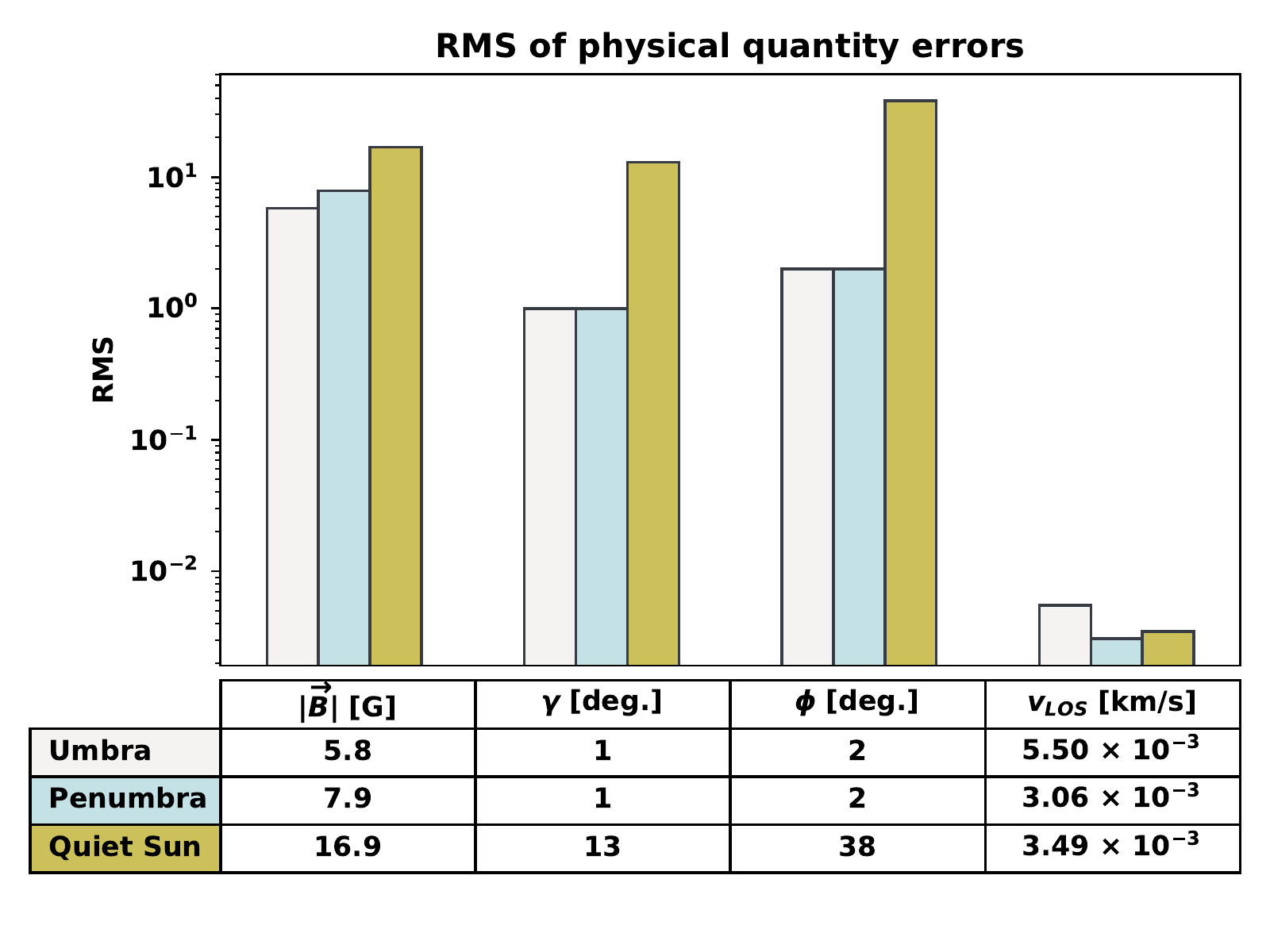}
      \caption{The RMS of the errors in the RTE inversion results primarily reflects the stability of the inversion in the different regions. It shows how the small errors in the input data, introduced through the processing, affect the final results retrieved with the same method. The error in $\lvert\vec B\lvert$ and $\gamma$ is smallest in the penumbra. The determination of $\gamma$ and $\phi$ is challenging in the quiet Sun, due to low signal levels, which is also reflected in the error RMS. $v_{\rm LOS}$ in the umbra has higher error due to a shallower line core and more complex profiles. (The magnitude of the errors is discussed in Sect.~\ref{sec:DisConclusion})}
         \label{Fig:RMS_PhyiscalParam}
   \end{figure}

The output parameters of the inversion are also configurable. Thus, we can request only a subset of the full set of output (physical) parameters for a given mode. For the first modes that calculate all nine Milne-Eddington model parameters, in standard operations we only request the four parameters of interest: the three components of the magnetic field vector and the LOS velocity. For the other modes in standard operations, we select all available outputs.
   
Before we perform the RTE inversion, we must prepare the data to match the interface of the inverter. After demodulation, the data are normalised to $I_c$ ($I_c$ is calculated on ground). The pipeline performs this operation with $I_c$ represented on 12.8 bits (which corresponds to it being calculated on data with maximum range $2^{12}$), and scales the result to a maximum range of $2^{23}$ . We transform these data from fixed to floating point by assuming a $2.30$ fixed point representation. As part of the preparation of the data, the pipeline also rearranges the images, to provide the inverter with a data stream that it can process (i.e. all 24 values that the same pixel in the FOV takes in the data set). A similar step takes place after the inversion, to form images again.

\begin{figure}[tbp]
   \centering
   
   \begin{minipage}[c]{\hsize}
    \centering
    
    \includegraphics[width=.30\hsize]{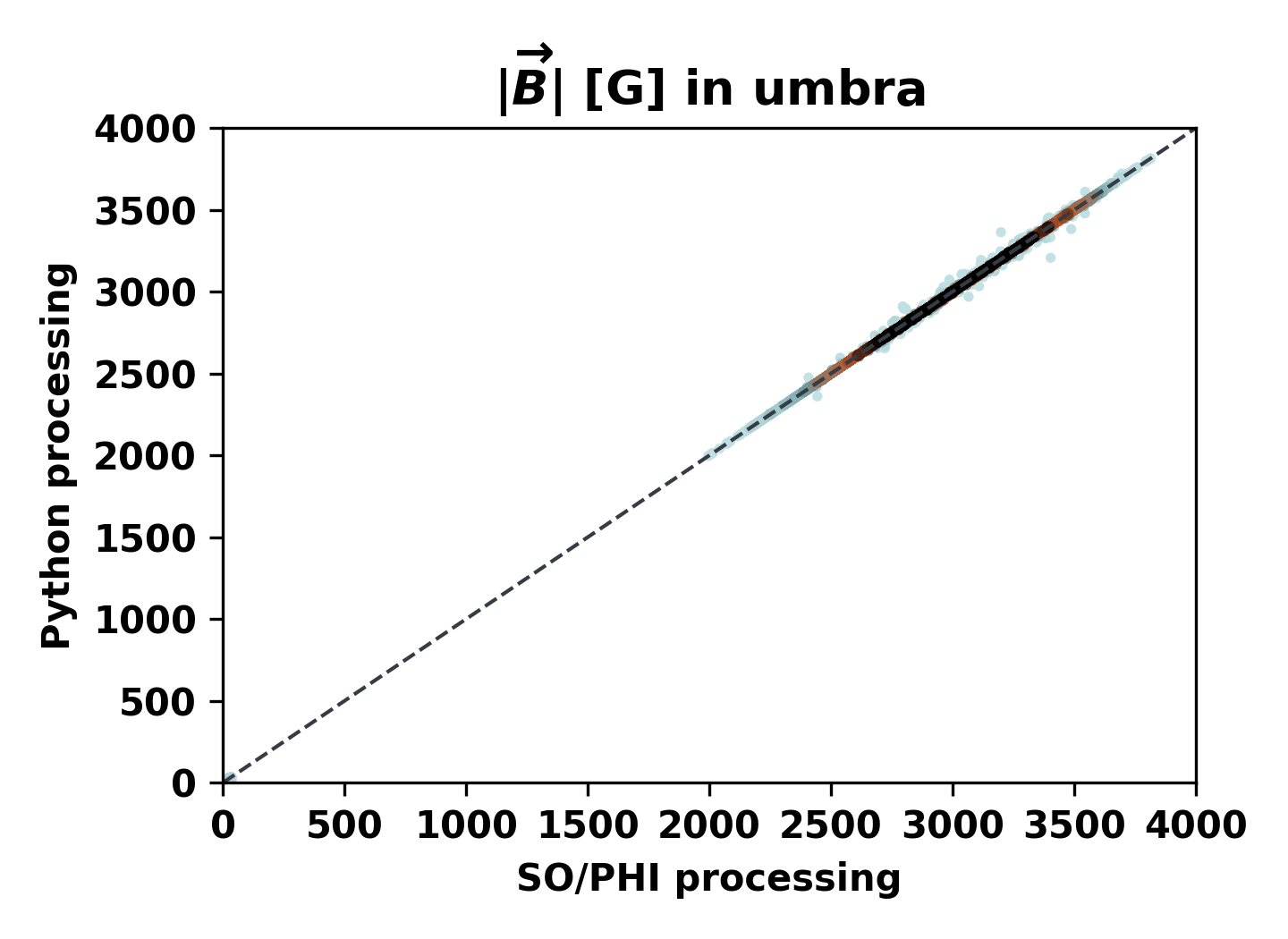}
    \includegraphics[width=.28\hsize]{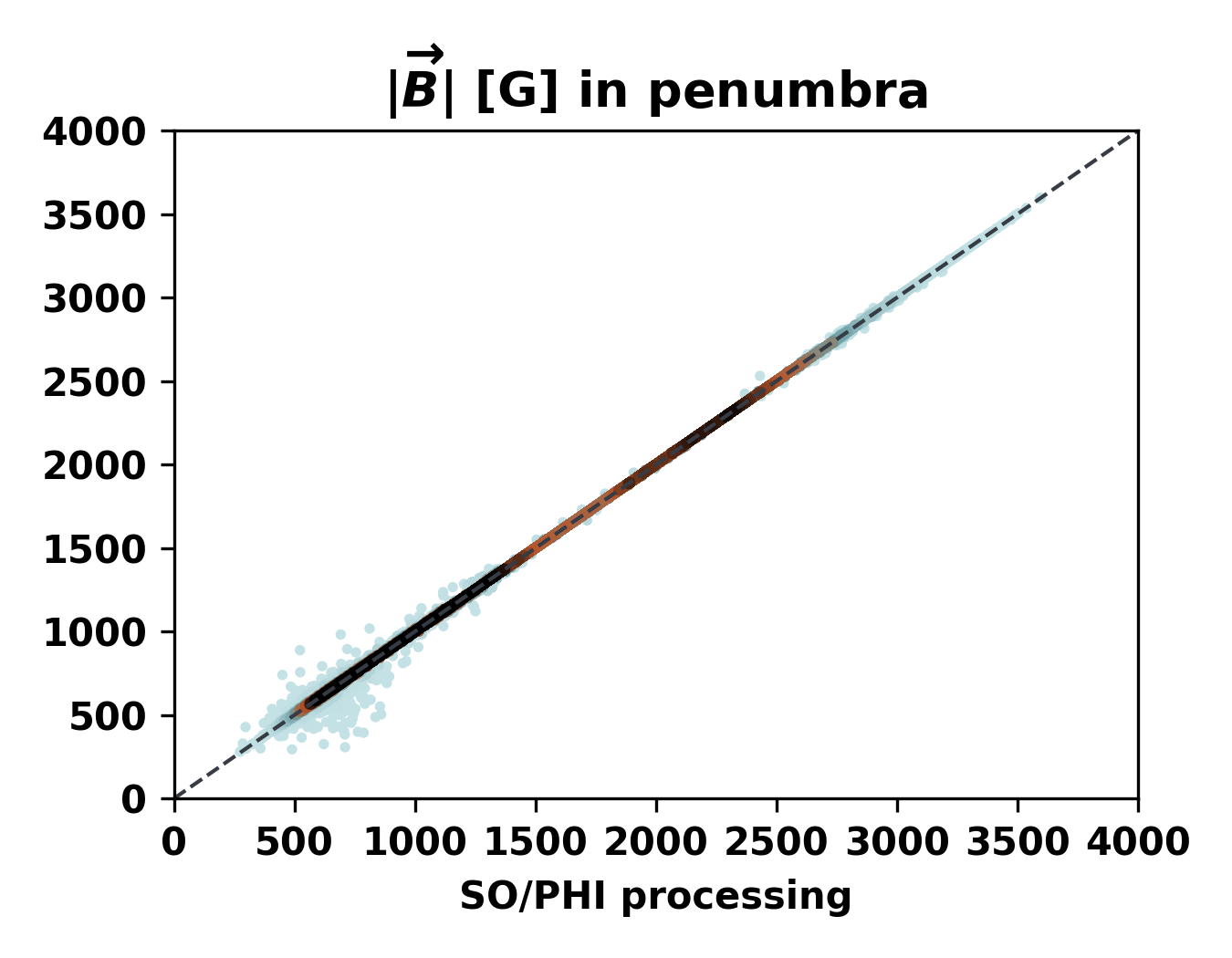}
    \includegraphics[width=.32\hsize]{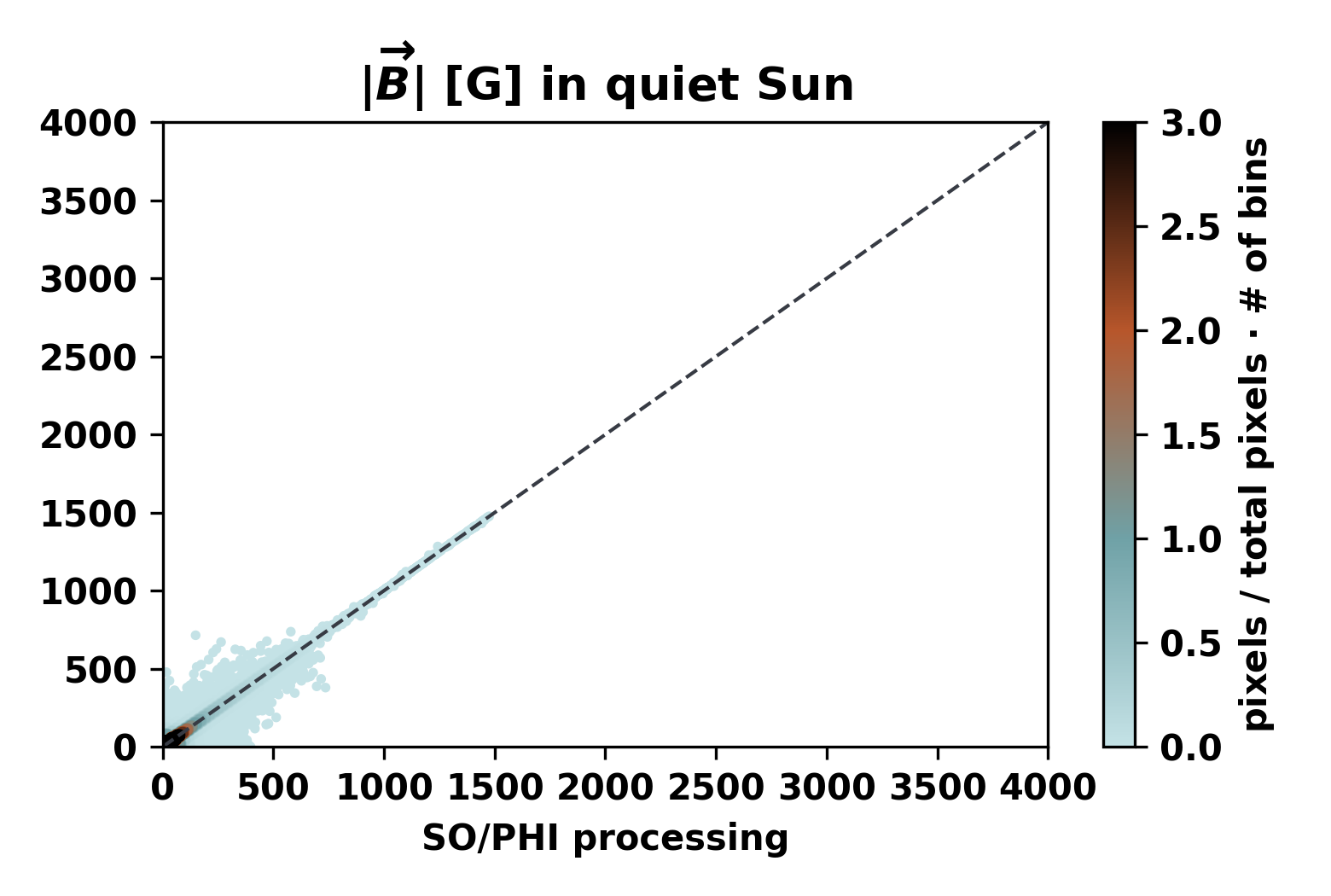}
    \end{minipage}
    
    \begin{minipage}[c]{\hsize}
    \centering
    
    \includegraphics[width=.30\hsize]{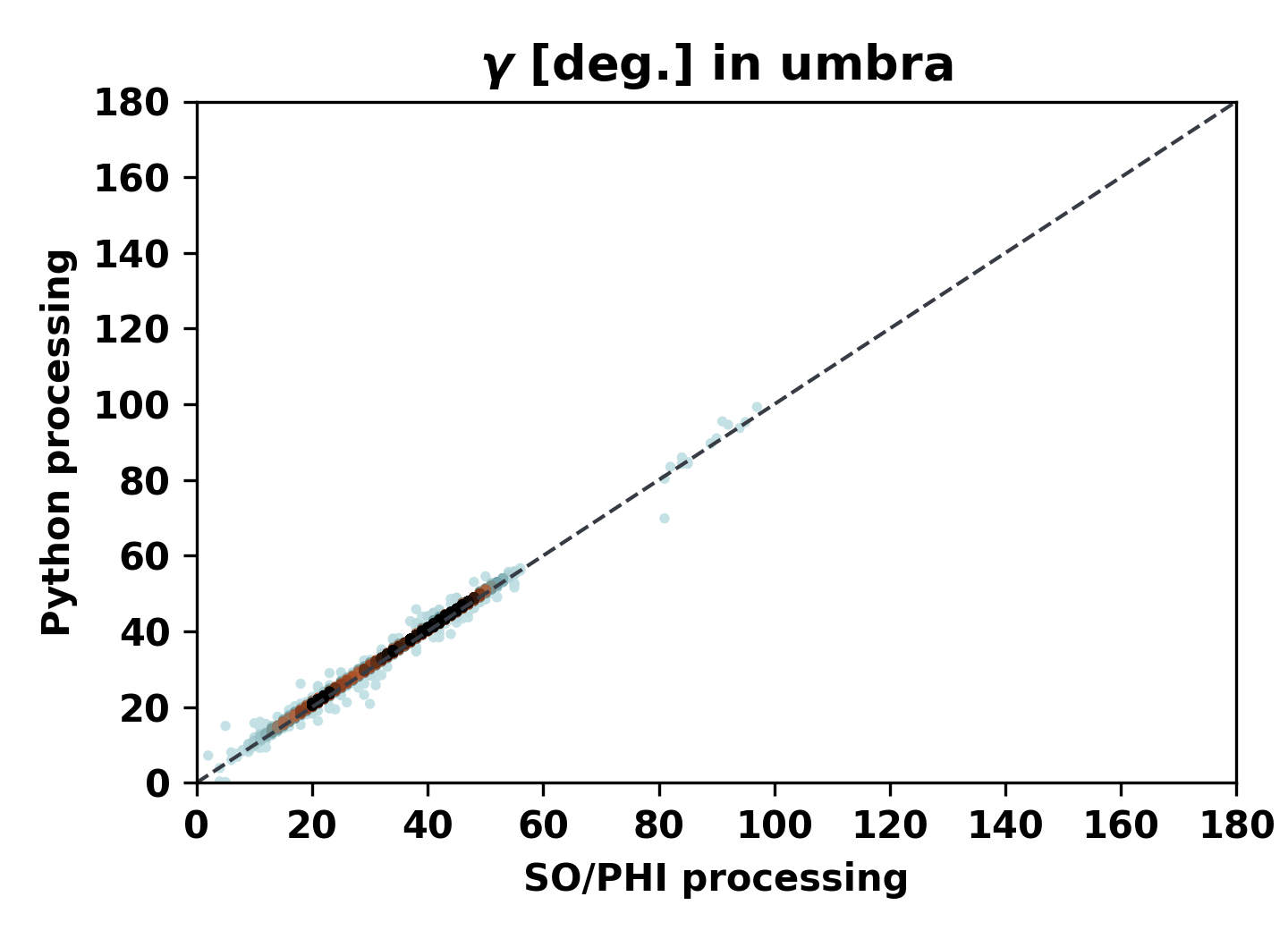}
    \includegraphics[width=.28\hsize]{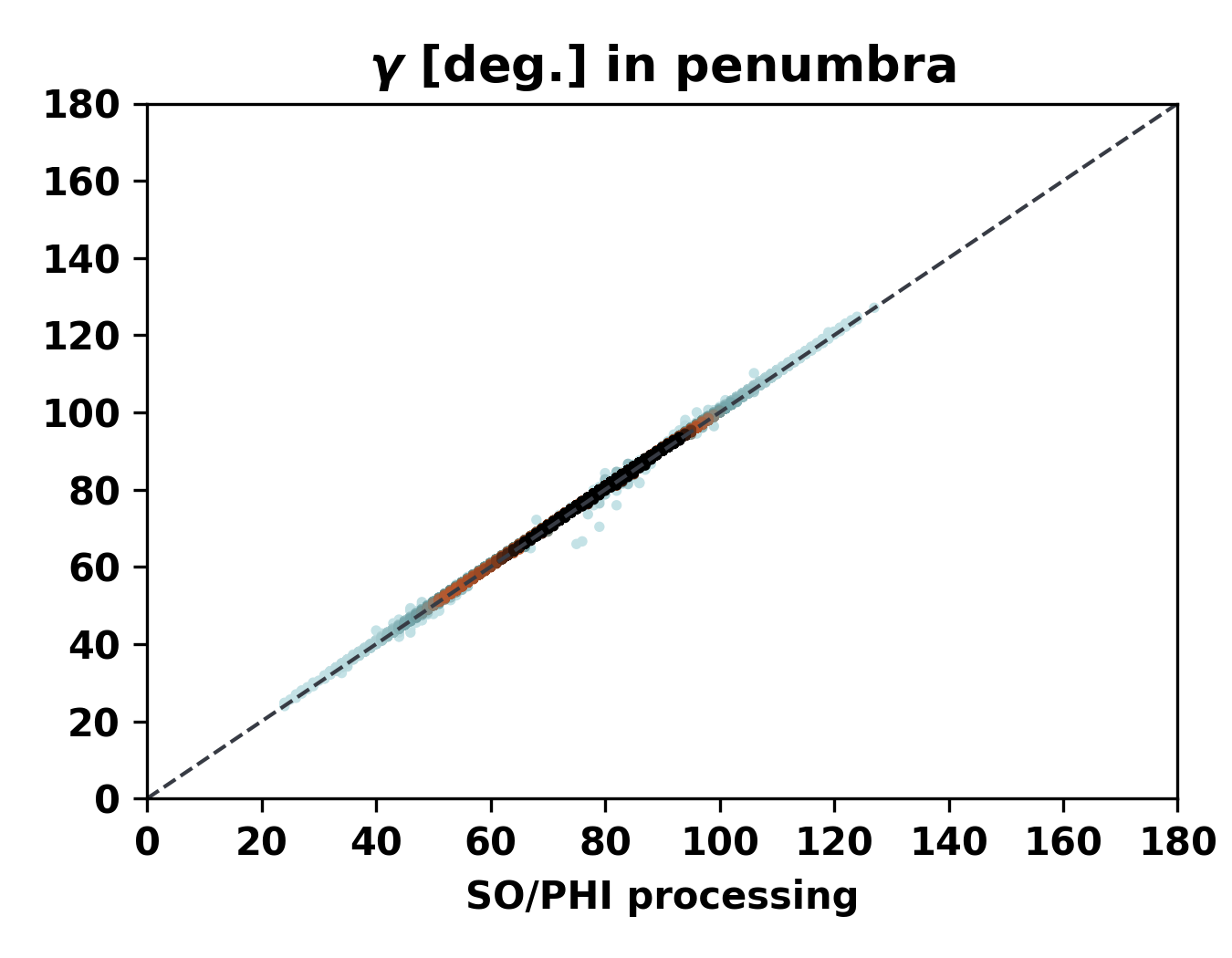}
    \includegraphics[width=.32\hsize]{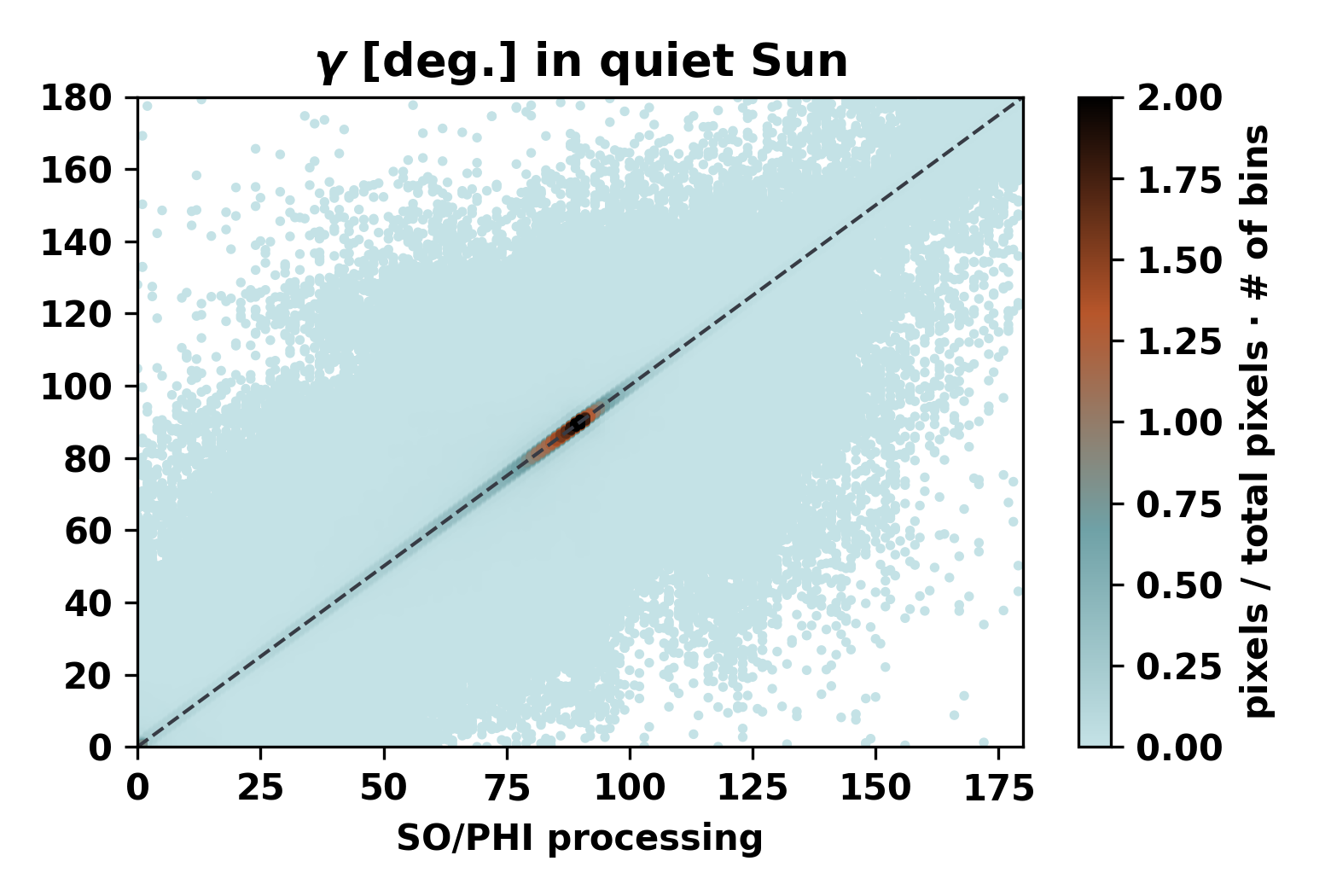}
    \end{minipage}
    
    \begin{minipage}[c]{\hsize}
    \centering
    
    \includegraphics[width=.30\hsize]{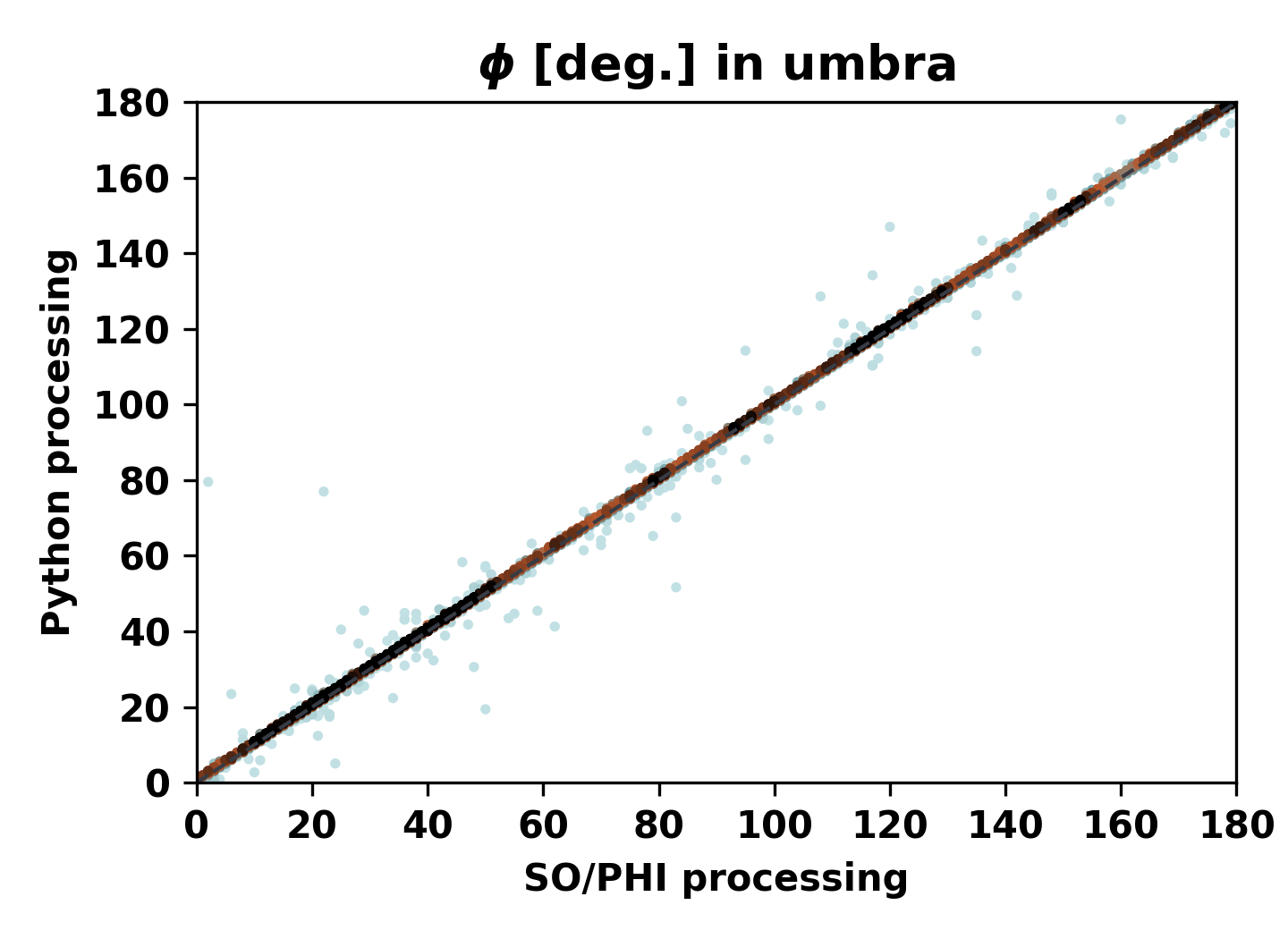}
    \includegraphics[width=.28\hsize]{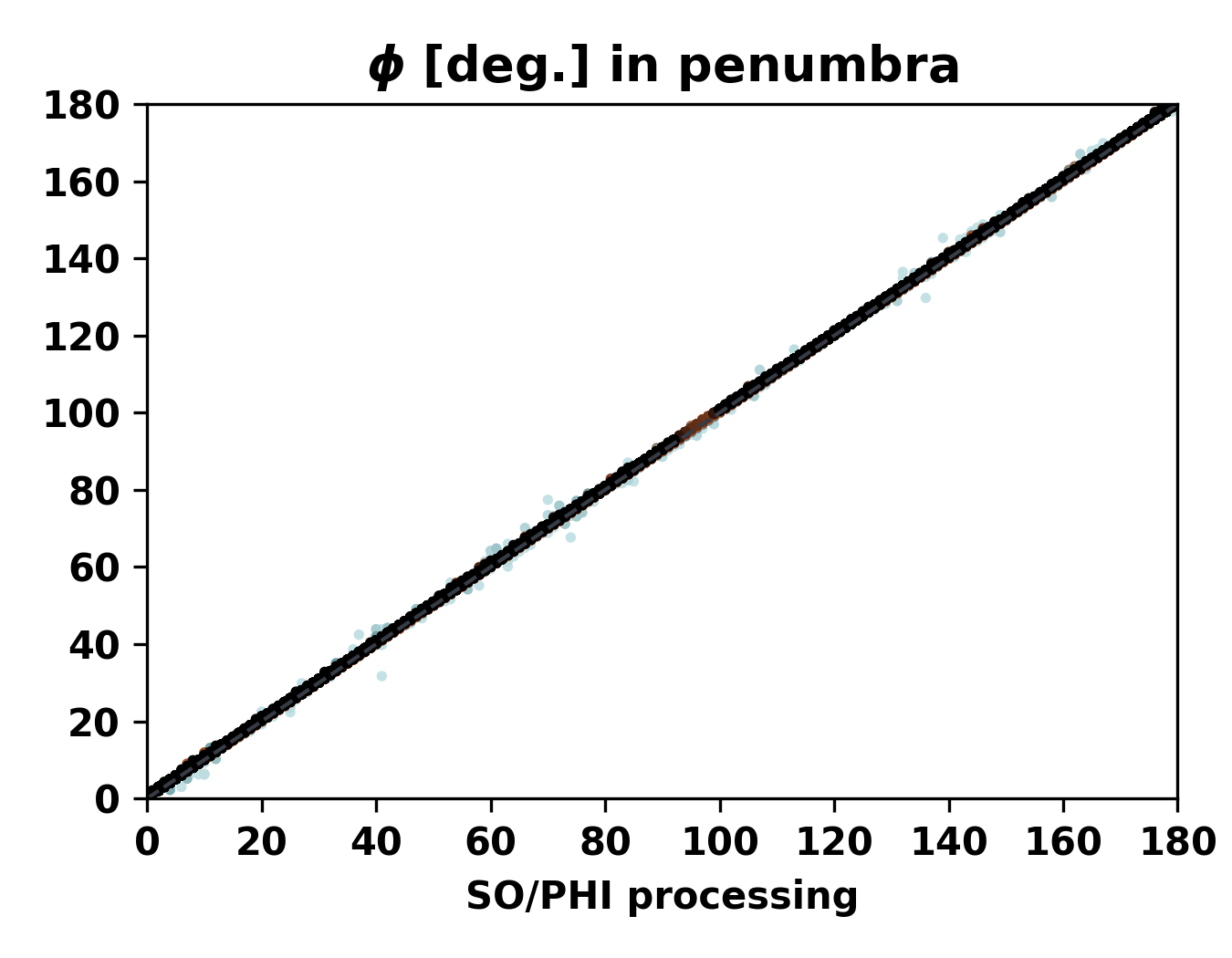}
    \includegraphics[width=.32\hsize]{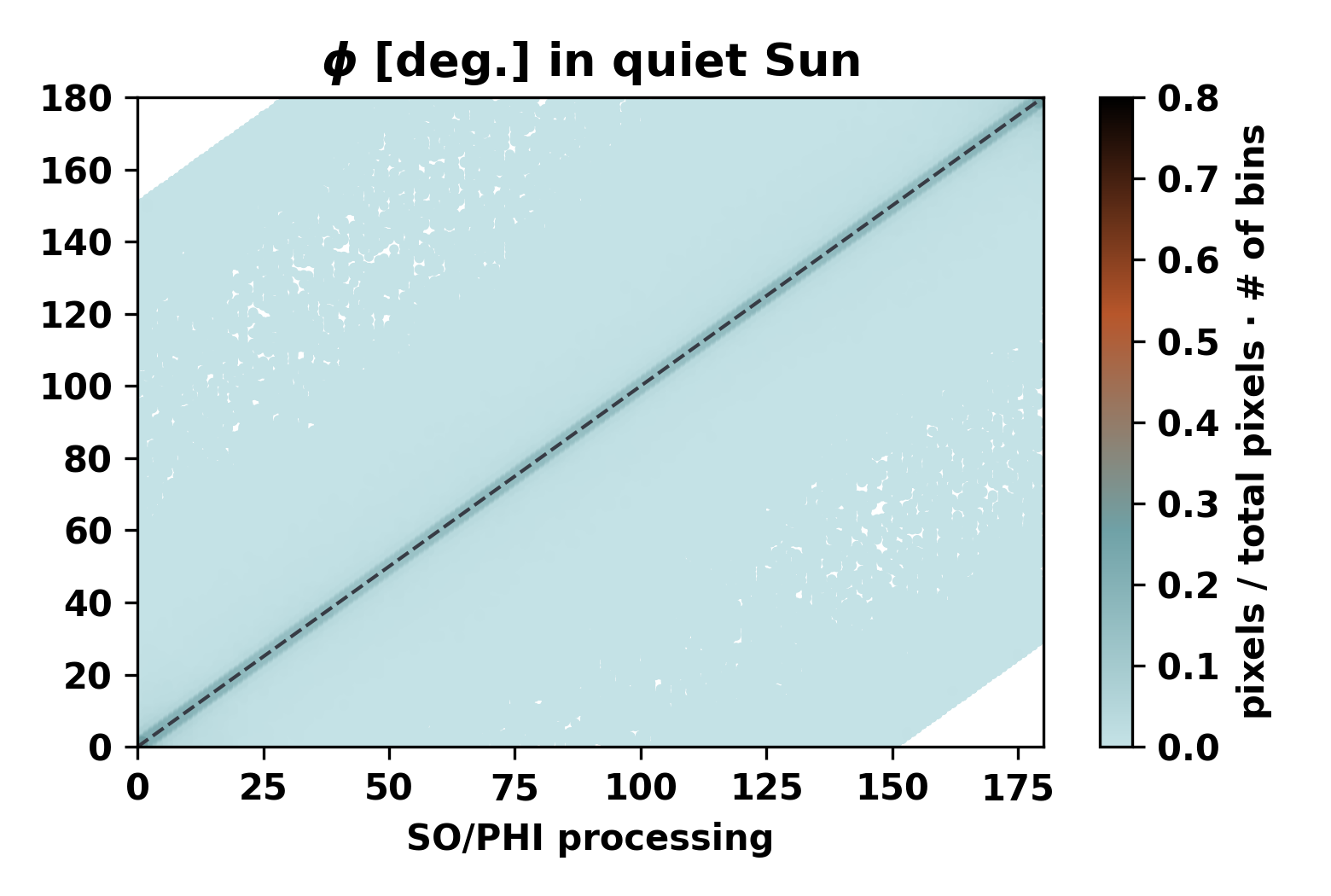}
    \end{minipage}

    \begin{minipage}[c]{\hsize}
    \centering
    
    \includegraphics[width=.30\hsize]{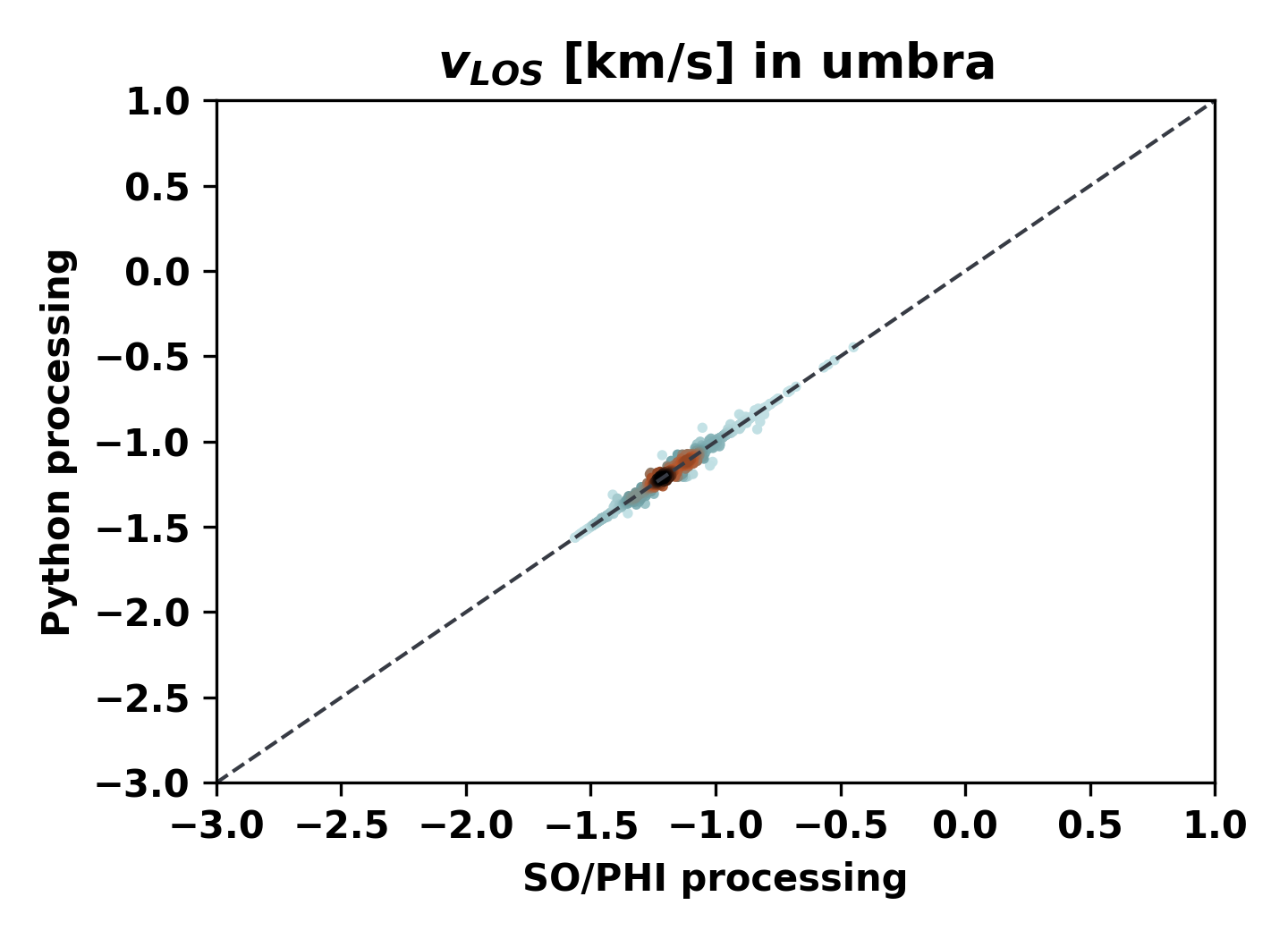}
    \includegraphics[width=.28\hsize]{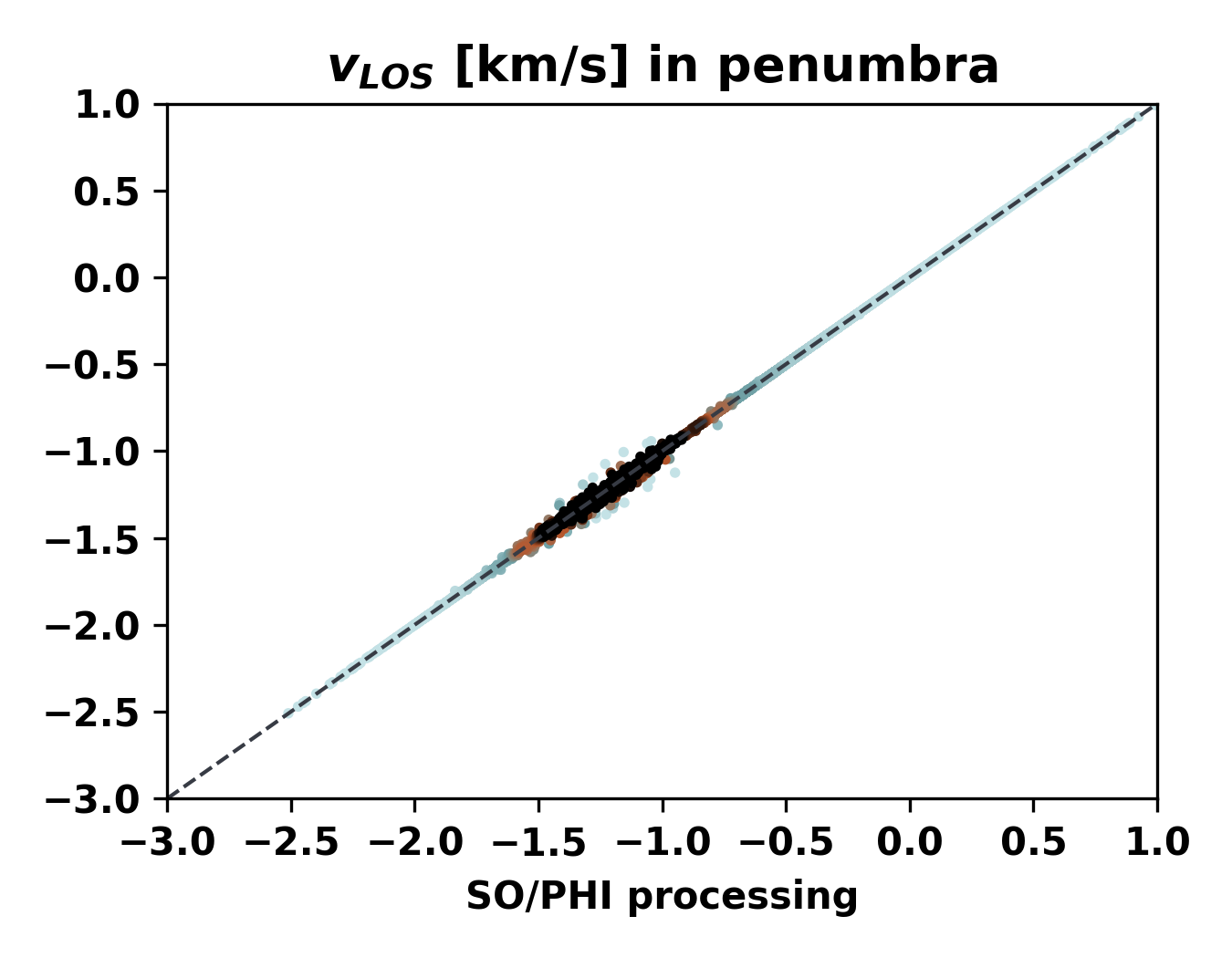}
    \includegraphics[width=.32\hsize]{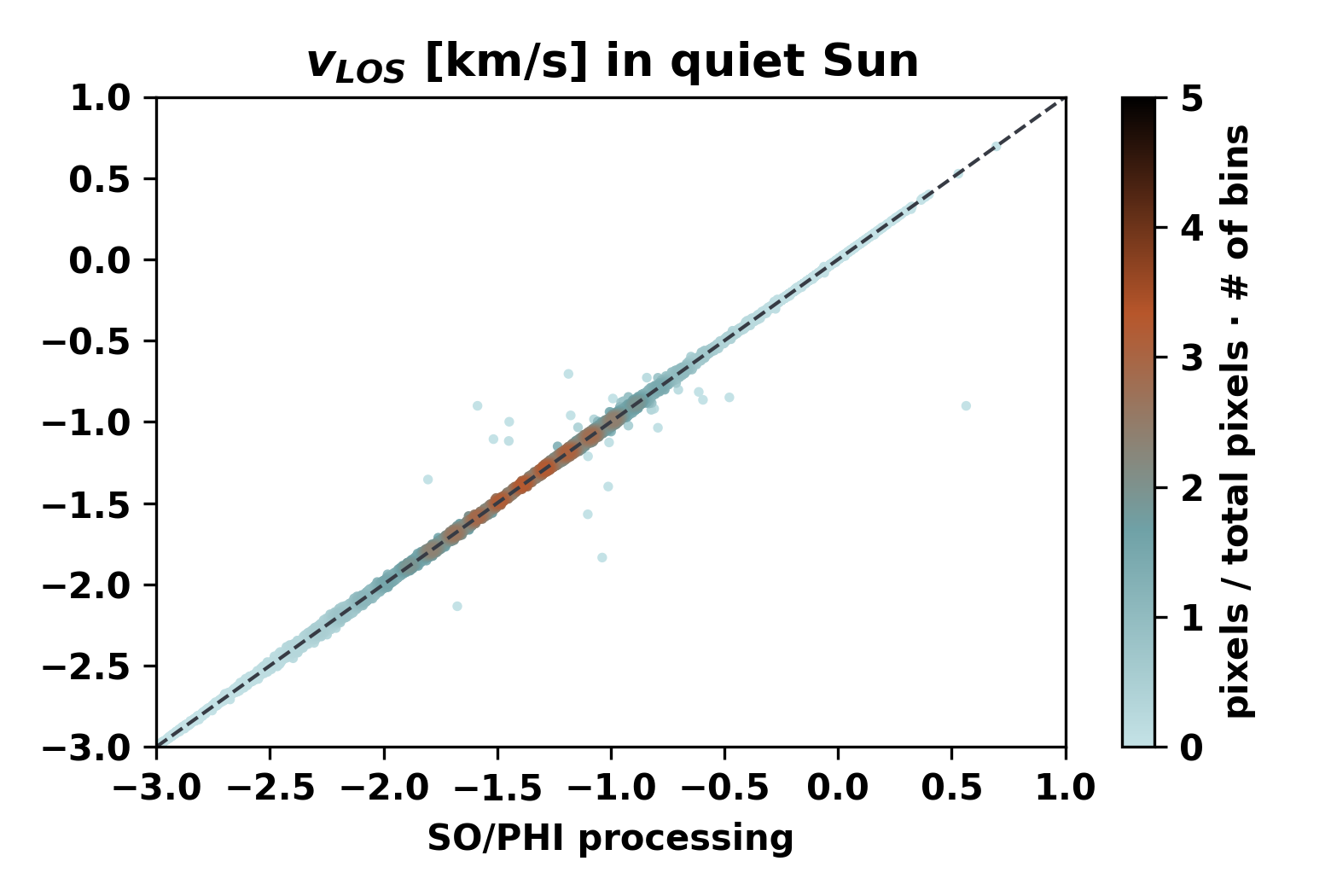}
    \end{minipage} 
    
     \caption{The correlation of the reference and \sophi{} processing results for $\lvert\vec B\lvert$, $\gamma$, $\phi$ and $v_{\rm LOS}$. While the Stokes vector errors are uniformly distributed in these regions (see Fig. \ref{Fig:BarPlotStokes}, the scatter of the results varies significantly. This is due to the different stability of the inversion in the different regions, discussed in the main text.}
         \label{Fig:Scatter}
   \end{figure}

We use the first mode of the inverter in this study, where we do the full inversion of the data, with initial conditions calculated through classical estimates. See Fig.~\ref{Fig:Results} for the results of the RTE inversion. These results show smooth transitions of the values over the FOV in $\lvert\vec B\lvert$, $\gamma$ and $v_{\rm LOS}$. In $\lvert\vec B\lvert$ we obtain magnetic fields up to $4000$\,G. The upper limit in the inversion module for $\lvert\vec B\lvert$ is $5000$\,G, which is not  reached. In $v_{\rm LOS}$ we can see the up- and downflows of the solar granulation, as well as the Evershed flow in the penumbra \citep[see][]{Evershed1909Radial}. The upper and lower limits of the inversion module for $v_{\rm LOS}$ is $[-20,20]$\,km/s. In $\phi$, the azimuth ambiguity disrupts the smooth transitions, and is dominated by noise in the quiet Sun.

The source of the differences that we obtain between the reference pipeline and the \sophi{} processing pipeline, are the small variations of the input data due to the processing. However, the stability of inversion on such a data set (considering the physics of the MHD simulation, the spectral and spatial convolutions applied, and the spectral sampling points) also determines the amount of error introduced by these small changes. In case of real \sophi{} observations, we would have an additional contribution from the inaccuracy of the calibration data. This is eliminated in this study by using the same data to degrade and calibrate the synthetic data set. In the following paragraphs, we present the results; we discuss their magnitude and significance in Sect.~\ref{sec:DisConclusion}.

The RMS of the errors in $\lvert\vec B\lvert$ introduced by the \sophi{} pipeline vary between $16.9$\,G to $5.8$\,G from the quiet Sun to umbra, see Fig.~\ref{Fig:RMS_PhyiscalParam}). Values below $1000$\,G show a larger disagreement as a consequence of lower signal levels, both in the case of the quiet Sun and the penumbra, showing up as a low density scatter in Fig.~\ref{Fig:Scatter}. We note, that pixels with values $1000$\,G belong to the outer penumbra, reaching the lower limit of typical field strengths in penumbral regions. A few outliers appear also in the umbra, particularly above $2500$\,G, hinting at the difficulty in inverting complex line profiles. These profiles, on one hand, are sampled by only six points, on the other hand they are also significantly changed by the prefilter and spatial PSF convolution (see Fig.~\ref{Fig:InputData_a}).

The errors in the orientation of the magnetic field vector we present to integer precision. This is due to the fact, that the RTE inversion truncates these values to integer accuracy, therefore any sub-decimal precision differences between the errors in the various regions would be an artefact of this operation. The umbral and penumbral regions show the same accuracy for both the inclination, $\gamma$, and the azimuth: $\phi$, $1$\,$^{\circ}$ and $2$\,$^{\circ}$, respectively. (see Fig.~\ref{Fig:RMS_PhyiscalParam}). In the quiet Sun, we see an error increase in both angles, a sign of lower signal strength. This results in noisy vector direction, with the correlation plot showing a large scatter across all possible values. The RMS of the errors in the quiet Sun regions is $13$\,$^{\circ}$ and $38$\,$^{\circ}$ for $\gamma$ and $\phi$, respectively. In the analysis of $\phi$, we assume that any difference between the two results larger than $150$\,$^{\circ}$ is caused by the ambiguity of the angle, and therefore we change all corresponding pixels to their supplementary angles. This results in the sharp cut at these errors, showing up as empty corners, in the last panel of Fig.~\ref{Fig:Scatter}.

$v_{\rm LOS}$ is calculated with the best precision in the penumbra and the quiet Sun, with an error RMS between $3.0$-$3.5$\,m/s (see Fig.~\ref{Fig:RMS_PhyiscalParam}). The increase of the accuracy in the penumbra is due to the stronger velocities in this region. The calculation of $v_{\rm LOS}$ in the umbra produces an error with $5.5$\,m/s RMS. These regions, with higher magnetic fields, produce stronger Zeeman splitting: the spectral profiles widen, become more shallow and complex. Due to our spectral sampling (as seen in Fig.~\ref{Fig:InputData_b}), the representation of these profiles becomes less accurate, and therefore, the sensitivity of the Stokes vector to the LOS velocity perturbations diminishes, leading to larger errors.

Using the vector magnetic field that we retrieve from the RTE inversion, we also calculate the line of sight magnetic field ($B_{\rm LOS}$), for further insight. The RMS error of the $B_{\rm LOS}$ is $16.9$\,G, $7.9$\,G and $5.8$\,G for the umbra, penumbra and quiet Sun, respectively. It shows the best agreement in the quiet Sun, significantly better than what $\lvert\vec B\lvert$ provided. This is due to the fact that the granules harbour weak transverse fields \citep[see][]{Orozco2012Internetwork, Danilovic2016internetwork}, which translate to small $B_{\rm LOS}$ values, lowering their contribution to the RMS of the error. In contrast, in the umbra and penumbra, we can observe an increase in the error. These regions harbour stronger magnetic fields, which appear at higher angles, all the way to close to vertical in the umbra. Their orientation and strength translates to strong $B_{\rm LOS}$ signals, creating a higher RMS error.
We note, that $B_{\rm LOS}$ can also be computed on-board without using an inversion, with analytical formulae (see above), which is a different approach, and these values can not be applied to it.

\section{Discussion and Conclusions}\label{sec:DisConclusion}

Our accuracy analysis shows that the on-board processing trade-offs do not compromise the accuracy of the \sophi{} data. The comparison between the results of the fully on-board processed data (with the necessary trade-offs) and the results obtained on-ground (without the trade-offs of the on-board processing) conveys that the errors in the determination of the final Stokes parameters are below $7 \times 10^{-5}$. This leaves a good margin for other sources of errors (e.g., calibration errors) to fulfil the $10^{-3}$ polarimetric sensitivity requirement of \sophi{}.

We present the errors in the physical parameters to give an idea of how the RTE inversion results may change by these processing inaccuracies. However, it is important to remark, that most input data sets intrinsically deviate from Milne-Eddington line profiles, and a tiny error in the input may cause the inversion to converge to a different local minimum, providing large differences in the physical parameters.

In \cite{albert2018performance} we have done a similar analysis on a data set acquired by the Helioseismic and Magnetic Imager on-board the Solar Dynamics Observatory \citep[SDO/HMI; ][]{Schou2012HMI}, using an earlier version of the \sophi{} data reduction pipeline. In that study, we compared the errors from the processing to a statistical analysis performed with the He-Line Information Extractor inversion code \citep[\helix{}; see][]{Lagg2004}. The statistical mode of \helix{} inverts the data with varying starting conditions, and we regard the variation of the results as a measure of the inaccuracy inherent in the inversion of the data set, amounting to $21.9$\,G, $1.34^\circ$, $1.37^\circ$, and $14.5$\,m/s for $\lvert\vec B\lvert$, $\gamma$, $\phi$ and $v_{\rm LOS}$, respectively. The errors in the inversion results, introduced to the SDO/HMI data by the on-board processing of \sophi{} \citep{albert2018performance}, are $33.64$\,G, $2.56^\circ$, $1.92^\circ$, and $19$\,m/s for the same parameters, which are slightly higher than what we find in the current study. This is due to the different input data, as well as the earlier version of the processing pipeline. It is important to note, that in the \cite{albert2018performance} study we determined the errors in the magnetic field vector limiting the FOV to an area with strong polarisation signals, while for the $v_{\rm LOS}$ we take the full FOV into consideration.

\cite{carrascosa2016rte}, while verifying the implementation of the RTE inverter on-board \sophi{}, compared inversions with the C and FPGA implementation of the code for a collection of Milne-Eddington synthetic profiles (considered as ideal input, containing only symmetric profiles). These profiles were sampled with $5$\,m\AA\  steps, and selected to range between $0$ and $1500$\,G in $\lvert\vec B\lvert$, $0$ and $180^\circ$ in $\gamma$ and $\phi$, and $-2$ and $2$\,m/s for $v_{\rm LOS}$, to which they added noise with a magnitude of $10^{-3} \times I_{c}$. The results agree with an error RMS of $5.3$\,G, $4.86^\circ$, $5.77^\circ$ and $5.9$\,m/s for $\lvert\vec B\lvert$, $\gamma$, $\phi$, and $v_{\rm LOS}$, respectively. However, the same comparison, performed on observations from the Swedish Solar Telescope, results in RMS errors of  $69.2$\,G, $6.5^\circ$, $5.47^\circ$, and $79.41$\,m/s. This is due to several factors, including higher noise, instrumental errors, and asymmetric data profiles, which is typical of observed solar Stokes profiles \citep[see e.g.][]{Sami1993Smallscale}. The data that we analyse in this paper fall between the two tests in \cite{carrascosa2016rte}: we do not introduce additional noise into our data, other than what the pipeline produces (which is in the order of $7 \times 10^{-5}$), however we do have asymmetry in some profiles. We find that the error introduced by the on-board processing pipeline is comparable to the error introduced by the inverter implementation when tested on synthetic data. The differences between the results on the different data sets indicate that the error resulting from the RTE inversion is dominated by the noise level of the data and the input data profiles, which is an inherent property of RTE inversions. This underlines the fact, that the accuracy of the pipeline can be judged best by comparing the Stokes parameters, instead of the results of the RTE inversion. In order to find a context for the errors of the physical parameters, retrieved by the RTE inversion, we must compare them to a very similar data set.

\cite{Borrero2014Comparison} compared different Milne-Eddington inversions using data from a sunspot simulation described in \cite{Rempel2012Numerical}, which is very similar to what we used in this work. They synthesised two absorption lines (\ion{Fe}{I}\,$630.15$\,nm and \ion{Fe}{I}\,$630.24$\,nm) over the whole MHD cube, and did not introduce any noise. The lines were sampled at 100 wavelength steps, $10$\,m\AA\ apart. This sampling provides more information than is available in \sophi{} observations, consequently a better reconstruction of the absorption lines is expected. Moreover, the authors cropped the field of view, such that it contains a comparable number of pixels in umbra, penumbra, and solar granulation (more precisely a $\sim200$\,G plage region surrounding the spot). Using this data set, they found that different Milne-Eddington inversion codes, executed on one input, produce values within an interval of $35$\,G, $1.2^\circ$ and $10$\,m/s for $\lvert\vec B\lvert$, $\gamma$, and $v_{\rm LOS}$, respectively. In this work, we find that the differences in the \sophi{} inversion of on-board and on ground reduced data are smaller than the differences introduced by different inversion codes executed on a single input in \cite{Borrero2014Comparison}.
This means, that the accuracy of the on-board processing is higher than the accuracy of a typical Milne-Eddington inversion. We furthermore note, that \cite{Borrero2014Comparison} does not discuss the errors introduced by Milne-Eddington inversions due to simplifications in the physics underlying this model. We expect these to be considerably larger than the numerical uncertainties between different Milne-Eddington codes \citep[see][and references therein]{Orozco2010Applicability, CastellanosDuran2022Phd}.

In conclusion, the \sophi{} pipeline provides the necessary accuracy to process spectropolarimetric data with $10^{-3}$ polarimetric sensitivity. We show, that the data processing pipeline does not compromise the accuracy of the inversion results, since it preserves the confidence interval of the Milne-Eddington RTE inversions. Comparing the results of this study with others shows that the effect of the on-board pipeline errors on the RTE inversion is below the errors produced by the RTE inversion inherently on both simulated and observed data. In this paper, we analyse the errors introduced by the on-board data processing pipeline in comparison to on-ground processing. However, these errors can be regarded as negligible, or at least small compared with other sources of error.

Similar processing accuracy can be expected in other on-board data processing pipelines as well (e.g., calculating flat fields, or determining polarimetric ad-hoc cross-talk correction terms), since the restrictions and solutions presented here overarch all on-board implementations. This study also shows, that while requiring a significant effort, on-board reduction of solar spectropolarimetric data is a viable option for future instruments, even with stringent limitations in computational resources. It significantly reduces telemetry requirements for \sophi{} (from 24 images, to five at most, in addition to obviating the necessity to download dark and flat fields), and will be particularly valuable for spectropolarimeters on deep-space missions.

%%% %%%%%%%%%%%%%%%%%%%%%%%%%%%%%%%%%%%%%%%%%%%%%%%%%%%%%%%%%%%
%% Bibliography
%
% Using BibTeX
%
%\bibliographystyle{spr-mp-sola}
%\bibliography{report}  
%
% Without BibTeX 
% \begin{thebibliography}{}
% \bibitem[\protect\citeauthoryear{Author}{Year}]{key}
%   <bibliographical entry>
%
% \bibitem[\protect\citeauthoryear{}{}]{}
%   
%  
% \end{thebibliography}

\backmatter

\bmhead{Acknowledgments}

We thank Amanda Romero Avila and Philipp L\"{o}schl for their work in synthesising the MHD cube for the input data. We use data provided by M. Rempel at the National Center for Atmospheric Research (NCAR). This work was carried out in the framework of the International Max Planck Research School (IMPRS) for Solar System Science at the Technical University of Braunschweig and the University of G\"ottingen. Solar Orbiter is a mission led by the European Space Agency (ESA) with contribution from National Aeronautics and Space Administration (NASA). The \sophi{} instrument is supported by the German Aerospace Center (DLR) through grants 50 OT 1201 and 50 OT 1901. The Spanish contribution is funded by AEI/MCIN/10.13039/501100011033/ (RTI2018-096886-C5, PID2021-125325OB-C5, PCI2022-135009-2) and ERDF “A way of making Europe”; “Center of Excellence Severo Ochoa” awards to IAA-CSIC (SEV-2017-0709, CEX2021-001131-S); and a Ramón y Cajal fellowship awarded to DOS. JSCD was also funded by the Deutscher Akademischer Austauschdienst (DAAD). The National Center for Atmospheric Research is sponsored by the National Science Foundation.

\begin{itemize}
\item Funding:
German Aerospace Center (DLR), 50 OT 1201, 50 OT 1901;
Spanish Research Agency, ESP2016-77548-C5, RTI2018-096886-B-C5;
European FEDER funds;
State Agency for Research of the Spanish MCIU, SEV-2017-0709;
Deutscher Akademischer Austauschdienst (DAAD).
\item Conflict of interest/Competing interests: Not applicable
\item Ethics approval: Not applicable
\item Consent to participate: Not applicable
\item Consent for publication: Not applicable
\item Availability of data and materials
\item Code availability: Not applicable
\item Authors' contributions: K.A. did the analysis, wrote the main manuscript text and prepared the figures. J.H., J.S.C.D. and D.O.S. provided advice during the analysis. All authors reviewed the manuscript.
\end{itemize}

\noindent
If any of the sections are not relevant to your manuscript, please include the heading and write `Not applicable' for that section.


\bibliography{sn-bibliography}% common bib file


\end{document}